\documentclass[11pt]{elsarticle}
\usepackage[margin=0.9in]{geometry}
\makeatletter
\def\ps@pprintTitle{%
 \let\@oddhead\@empty
 \let\@evenhead\@empty
 \def\@oddfoot{}%
 \let\@evenfoot\@oddfoot}
\makeatother
\newcommand{\update}[1]{{\color{red} #1}}
\usepackage[ruled,vlined]{algorithm2e}
\usepackage[toc]{appendix}
\usepackage{rotating}

\usepackage{booktabs}
\usepackage[T1]{fontenc}
\usepackage{babel}
\usepackage[toc]{appendix}
\usepackage{lineno}
\usepackage{caption} 
\usepackage{setspace}
\usepackage{bm}%for bold
\usepackage{amsmath,amssymb,amsthm,bm}
\usepackage{graphicx}
\usepackage{multicol}
\usepackage{flafter}
\usepackage{color,hyperref,xcolor}
\usepackage{bbm}
\hypersetup{colorlinks=true,urlcolor=blue,citecolor=purple}
\usepackage{cleveref}
\usepackage{subcaption}
\renewcommand{\update}[1]{{\color{red} #1}}
\newtheoremstyle{general}
{3mm} % Space above
{3mm} % Space below
{} % Body font
{} % Indent amount
{\bfseries} % Theorem head font
{.} % Punctuation after theorem head
{.5em} % Space after theorem head
{} % Theorem head spec (can be left empty, meaning `normal')

\theoremstyle{general}

\usepackage{booktabs}
\numberwithin{equation}{section}

\def\U{\mathcal{U}}

\def\gauss{\mathcal{N}}

\newcommand{\abs}[1]{\left\lvert #1 \right\rvert}
\newcommand{\norm}[1]{\left\| #1 \right\|}

\DeclareMathOperator*{\argmin}{arg\,min}

\allowdisplaybreaks

\usepackage{setspace}

\makeatletter
\renewcommand{\fnum@figure}{Fig. \thefigure}
\makeatother
\biboptions{authoryear}

\begin{document}

\begin{frontmatter}

\title{A divide-and-conquer approach for spatio-temporal analysis of large house price data from Greater London}

%% Group authors per affiliation:
\author{Kapil Gupta, Soudeep Deb}
\address{Indian Institute of Management Bangalore \\ Bannerghatta Main Road, Bangalore, Karnataka 560076, India.}

%% or include affiliations in footnotes:

\begin{abstract}
% Statistical research in real estate markets, particularly understanding spatio-temporal dynamics of house prices, has garnered attention in recent times. Albeit Bayesian methods are common in spatio-temporal modeling, standard Markov chain Monte Carlo (MCMC) techniques are usually slow for large datasets such as the house price data. We propose a divide-and-conquer spatio-temporal modeling approach to tackle this problem. The method involves partitioning the data into multiple subsets and utilizing an appropriate Gaussian process model for each subset in parallel. The results from each subset are then combined via the Wasserstein barycenter technique to obtain the global parameters for the original problem. The divide-and-conquer approach allows multiple observations per spatial and time unit, thereby offering added benefit for practitioners. As a real life application, we analyze house price data from 983 middle layer super output areas in London for a period of eight years. The methodology renders insightful findings about the effects of various amenities, trend pattern, and relationship of price to carbon emission. Further, as demonstrated from a cross-validation study, it records good predictive accuracy while balancing the computational need.

Statistical research in real estate markets, particularly in understanding the spatio-temporal dynamics of house prices, has garnered significant attention in recent times. Although Bayesian methods are common in spatio-temporal modeling, standard Markov chain Monte Carlo (MCMC) techniques are usually slow for large datasets such as house price data. To tackle this problem, we propose a divide-and-conquer spatio-temporal modeling approach. This method involves partitioning the data into multiple subsets and applying an appropriate Gaussian process model to each subset in parallel. The results from each subset are then combined using the Wasserstein barycenter technique to obtain the global parameters for the original problem. The proposed methodology allows for multiple observations per spatial and time unit, thereby offering added benefits for practitioners. As a real-life application, we analyze house price data of more than 0.6 million transactions from 983 middle layer super output areas in London over a period of eight years. The methodology provides insightful findings about the effects of various amenities, trend patterns, and the relationship between prices and carbon emissions. Furthermore, as demonstrated through a cross-validation study, it shows good predictive accuracy while balancing computational efficiency.

% \sd{Abstract in 100 words (might be needed for JRSSA) suggested by ChatGPT:} Recent statistical research in real estate markets focuses on the spatio-temporal dynamics of house prices. Bayesian methods are common but often slow for large datasets like house price data. We propose a divide-and-conquer approach, partitioning the data into subsets and applying Gaussian process models in parallel. The results are combined using the Wasserstein barycenter technique to obtain global parameters. This method allows for multiple observations per spatial and time unit. Analyzing house price data from London, the methodology reveals insights into amenities, trends, and price-carbon emission relationships, demonstrating good predictive accuracy and computational efficiency.
\end{abstract}

\end{frontmatter}

%\linenumbers

\section{Introduction}
\label{sec:introduction}

The valuation of residential properties is immensely important for various economic stakeholders, including property dealers, sellers, and buyers. Traditionally, house price modeling has revolved around the hedonic framework \citep{rosen1974hedonic}. Specifically, log-linear regression techniques have been at the forefront in the analysis of house price transactions \citep{liu2013spatial}. However, the incorporation of location attributes into the mean structure of hedonic regression, as recommended by \cite{dubin1990specification}, has shown limitations in capturing all significant neighborhood attributes, potentially giving rise to spatial autocorrelation concerns. To overcome this challenge, researchers have advocated for the inclusion of spatial effects into the error structure \citep{can1990measurement, pace1998generalizing}. Simultaneously, the dynamics of house prices are profoundly influenced by temporal autocorrelation as well. \cite{pace1998spatiotemporal} leveraged the concepts of spatio-temporal modeling to comprehensively capture both spatial and temporal effects within house price data. This modeling approach has since become a cornerstone in academic research. Building on this foundation, \cite{liu2013spatial} harnessed the spatio-temporal autoregressive (STAR) model to address correlated errors in traditional hedonic regression, delving into spatial and temporal dependence considerations and further developing a house price index as a pivotal part of their analysis. Interestingly, their methodology assumed the spatial weight matrix to be lower triangular, which implies that two spatial units cannot mutually influence each other. A different approach was taken by \cite{fotheringham2003geographically} who integrated geographically weighted regression with time series forecasting techniques to capture the spatio-temporal variations in house prices. This was later improved with spatio-temporal data mining techniques by \cite{soltani2021housing}. \cite{holly2010spatio}, on the other hand, developed a spatio-temporal econometric model for analyzing house price dynamics in the United States, by explicitly taking into  account both cross-sectional dependence and heterogeneity. We also find it prudent to refer to the early review of spatial and spatio-temporal models within housing literature by \cite{gelfand2004dynamics}. The authors scrutinized several hierarchical models and briefly explored their Bayesian implementation to demonstrate the effectiveness of a Bayesian framework in this domain. Intriguingly, except for few studies (e.g., Bayesian implementation of STAR by \cite{beamonte2010analysis}, Bayesian network approach by \cite{teye2017detecting}), Bayesian spatio-temporal models have not been utilized appropriately in related statistical analysis. We attempt to bridge this gap in this article by proposing an effective Bayesian spatio-temporal approach in conjunction with a divide-and-conquer technique, to analyze large house price datasets.

It is of the essence here to briefly review a few motivating literature. The advent of the big data era has ushered in a new set of challenges for statistical inference, particularly in the realm of current topic of discussion. Complex Gaussian process models, though flexible, become computationally unwieldy for large datasets that feature numerous spatial locations and time-points, demanding substantial computational and storage resources. Various methods have been proposed to mitigate these challenges. A few examples are covariance tapering \citep{furrer2006covariance}, low-rank approximation \citep[][]{banerjee2008gaussian, wikle2010low}, and nearest-neighbor Gaussian process models \citep[][]{datta2016hierarchical, finley2019efficient}. On the other hand, the limitations of conventional Markov Chain Monte Carlo (MCMC) algorithms in dealing with vast datasets prompted the development of few scalable algorithms, including sub-sampling-based approaches \citep{quiroz2019speeding} and stochastic gradient MCMC \citep{nemeth2021stochastic}. Nevertheless, these methods often introduce additional errors and require extensive theoretical analysis to ensure their validity. An alternative solution in a Bayesian framework was proposed by \cite{guhaniyogi2018meta} and \cite{guhaniyogi2022distributed} which work as our primary motivation in this study. The key idea here is to employ a divide-and-conquer (hereafter abbreviated as D\&C) strategy for modeling Gaussian processes on massive spatial datasets.

The effectiveness of D\&C algorithms, as evident both in practical applications and theoretical foundations, has spurred our exploration of its potential within the realm of real estate markets. In this paper, we expand the D\&C approach into the spatio-temporal domain, dubbing it as the ``Divide-and-Conquer method for Spatio-Temporal Big Datasets'' (D\&C-STBD), and implement it to comprehensively analyze the house price dynamics of London. The proposed methodology entails data segmentation, parallel MCMC inference for each segment, and subsequent merging of the segment posteriors. To facilitate the last step, we employ the Wasserstein barycenter-based approach detailed in \cite{shyamalkumar2022algorithm}. It should be noted that our proposed method allows multiple (possibly unequal number of) and missing (zero) observations for every space-time combination, thereby obviating the need to aggregate individual observations within every spatio-temporal unit. Prior studies in spatio-temporal modeling have predominantly revolved around a balanced design regarding the number of observations at specific locations and time-points, cf.\ \cite{banerjee2014hierarchical}, \cite{sahu2022bayesian}. However, there has been limited research that incorporates unequal number of observations per spatial and temporal unit. 

As we show in our application, by embracing individual-level data (i.e., property-specific information as opposed to aggregated information per region), we can sidestep the need for aggregating covariate values into a generalized spatio-temporal structure. Subsequently, we are able to obtain more detailed insights about the effects of property characteristics and regional effects, as well as spatial and temporal dependence in determining house prices. Our analysis also shows that the model is adaptable and capable of predicting property prices, even for locations not observed in the training data. Several interesting aspects, such as the dependence of house price on carbon emission levels, effects of specific types of amenities, and the trend patterns are explored in detail. It is imperative to mention that while several studies have analysed UK house prices \citep[see, e.g.,][]{ahlfeldt2014form, feng2016postcode, cook2016new, chi2021shedding, mete2022hybrid, blatt2023changepoint}, to the best of our knowledge, there has not been an attempt to effectively address both spatial and temporal phenomena and dependence in this regard. 

Rest of this paper is organized as follows. In \Cref{sec:data}, we introduce the data, along with pertinent exploratory analysis. \Cref{sec:sp-model} present our proposed D\&C-STBD method, discussing its implementation and evaluation in \Cref{sec:model_matrices}. The main application is illustrated in \Cref{sec:results}. We conclude with some important remarks and future scopes of research in \Cref{sec:conclusion}. Technical details are deferred to the supplementary material.

\section{Data}
\label{sec:data}

\subsection{Data preparation}
\label{sec:data-prep}

In regional science and urban research, open government data is increasingly being used for spatial analysis of urban areas, with a growing emphasis on data accessibility \citep{arribas2014accidental}. In the United Kingdom, HM Land Registry provides open access to Price Paid Data\footnote{Price Paid Dataset URL: \url{https://www.gov.uk/government/statistical-data-sets/price-paid-data-downloads}}(PPD) detailing property sales transactions in England and Wales, including sale records and basic physical attributes.
Another important resource is the Energy Performance Certificates (EPC) dataset\footnote{The Energy Performance Certificates dataset URL: \url{https://epc.opendatacommunities.org/docs/guidance}}, offering insights into attributes like floor area, energy ratings, carbon emissions, and heating costs of buildings in England and Wales. Although the PPD dataset contains valuable information, it lacks essential property attributes such as total floor area. Without these details, the analysis of house price data is incomplete. To address this gap, many researchers \citep[e.g.,][]{chi2021new} have combined PPD and EPC datasets to analyze property transactions from 2011 to 2019. This combined dataset includes various geospatial attributes like postcodes, Output Areas (OAs), Lower-layer Super Output Areas (LSOAs), Middle-layer Super Output Areas (MSOAs), and districts for each location. We use the same idea and combine the two datasets in the current analysis. 

It is critical to point out that the OAs serve as the foundational elements for spatial census statistics, deliberately structured to ensure a certain level of socioeconomic coherence. LSOAs, a specialized form of census geography, typically encompass an average of five OAs each. In turn, MSOAs combine approximately five LSOAs on average and are situated within district boundaries. These carefully defined geographic units are meticulously crafted by the Office for National Statistics (ONS) specifically for analytical purposes. They are often used as the spatial units, as demanded in specific research studies, to examine the spatial variability and socioeconomic trends of key variables. For the house price data, frequent transactions are not observed at many OA and LSOA levels, resulting in a significant number of missing values in spatio-temporal settings. Consequently, interpreting data at these levels may not yield robust insights. Conversely, the MSOA level exhibits a substantial number of transactions, making it a more viable option for spatial resolution in our study. Therefore, we opt to utilize the MSOA level for our spatial analysis, which offers a more comprehensive understanding of the data landscape. The analysis of house prices at the MSOA level has been previously investigated in the literature (see \cite{chi2021shedding}, \cite{chi2022delineating}).

We however faced the challenge that latitude and longitude information, crucial for spatial analysis, are not available in these datasets. This is resolved by utilizing \cite{freemap} as it helps us link the spatial units with latitude and longitude. We utilize the centroid of the geographical coordinates of all properties within each MSOA as the geographical coordinates for that MSOA. For the temporal resolution, we opt for a monthly time scale, which allows for a comprehensive analysis of monthly trends across all MSOAs. Furthermore, there are (potentially) multiple or no property transactions for different MSOAs at various time-points. We address both aspects within our modeling framework, and they are taken care of appropriately in the main analysis. The methodology will be elaborated in the following section. Overall, this study specifically centers on the London region, resulting in 983 MSOAs spanning 106 time-points (approximately 8.5 years of monthly data). After excluding the newly built properties that represent <1\% of the observations and do not have relevant information, we get 104,198 space-time combinations, with a total of 651,202 property transactions. For ease of understanding, in \Cref{tab:variables}, we summarize the variables considered in this study, along with their descriptions.  

 \begin{table}[!htb]
	\centering
	\caption{Variables considered in the main analysis and their descriptions.}
	\label{tab:variables}
	\begin{tabular}{p{2.2cm}p{2.5cm}p{2.2cm}p{7.2cm}}
		\hline
        \hline
		Category & Variable & Type & Description \\
		\hline
        Spatial unit & MSOA & Coordinates & Longitude and latitude of a given MSOA. \\
        Temporal unit & Monthly & Discrete & Month in which the sale was completed.  \\
        \hline 
        Response & Price & Numeric & House price per square meter.  \\	
		  \hline
        Covariates & Area & Numeric & Total floor area (in square meters). \\
		& Rooms & Categorical & Number of rooms in the property, categorized in 5 levels: cat1 ($\leqslant 2$ rooms), cat2 (3 rooms), cat3 (4 rooms), cat4 (5 rooms), cat5 ($>5$ rooms). \\
		& Property type & Categorical & Type of property, categorized in 4 levels: Detached, Semi-detached, Terraced, Flats. \\
		& Ventilation & Categorical & Type of mechanical ventilation in property, categorized in 3 levels: Natural, Extract-only, Both supply and extract. \\
		& Fireplace & Binary & 1 if there is fireplace in property, 0 otherwise.  \\
        & Wind turbine & Binary & 1 if there is a wind turbine, 0 otherwise. \\
		& CO2 emission & Numeric & CO2 emissions per year (in $kg/m^2$). \\
        \hline
        \hline
	\end{tabular}
\end{table}

\subsection{Exploratory analysis}
\label{sec:exp_analysis}

\begin{figure}[!htb]
    \centering
    \includegraphics[width = 0.7\textwidth,keepaspectratio]{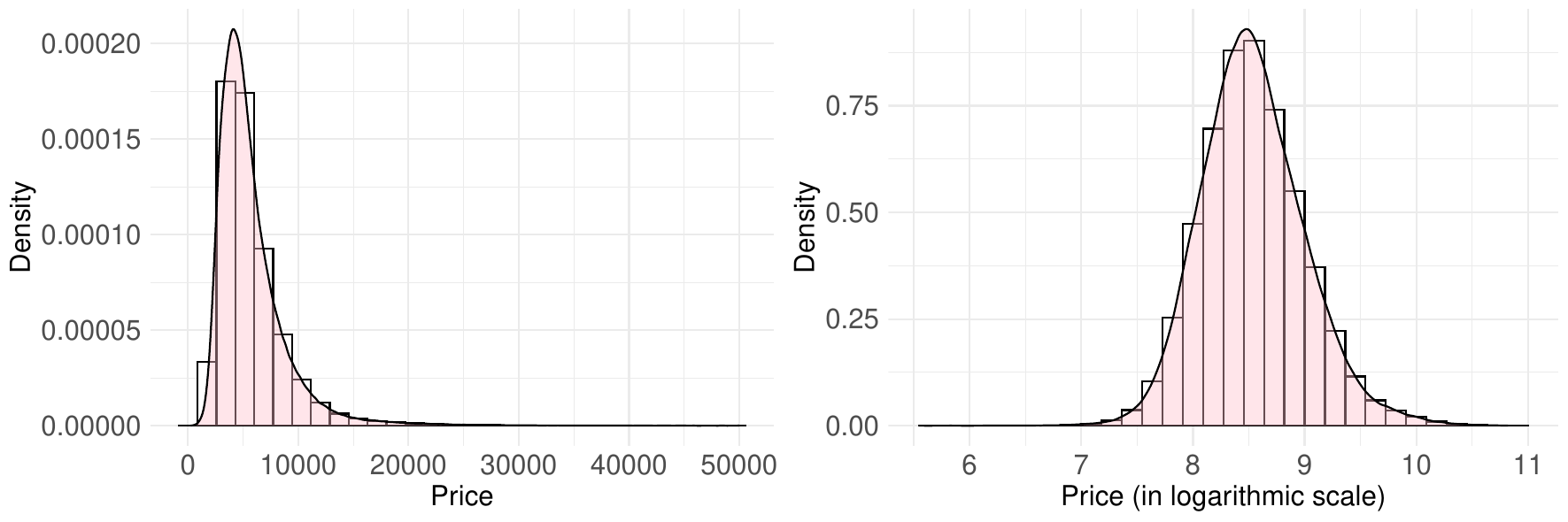}
    \caption{Density plot on histogram for the distribution of price per square meter for all properties in the data.}
    \label{fig:price}
\end{figure}

Let us start with a visual depiction of the distribution of the key variable in this study, the price per square meter of all transactions. In the left panel of \Cref{fig:price}, we  present the histogram and density plot for this variable. The distribution is clearly right-skewed, which is a common observation in real estate data \citep{curto2015listing}. This non-Gaussianity motivates us to adopt a logarithmic transformation, rendering the data normal, as evident in the right panel of the figure. It must be noted that the transformation log(price per square meter) is utilized in all further explorations in this article. 

\begin{figure}[!htb]
    \centering
    \vspace{-2cm}
    \includegraphics[width = 0.7\textwidth,keepaspectratio]{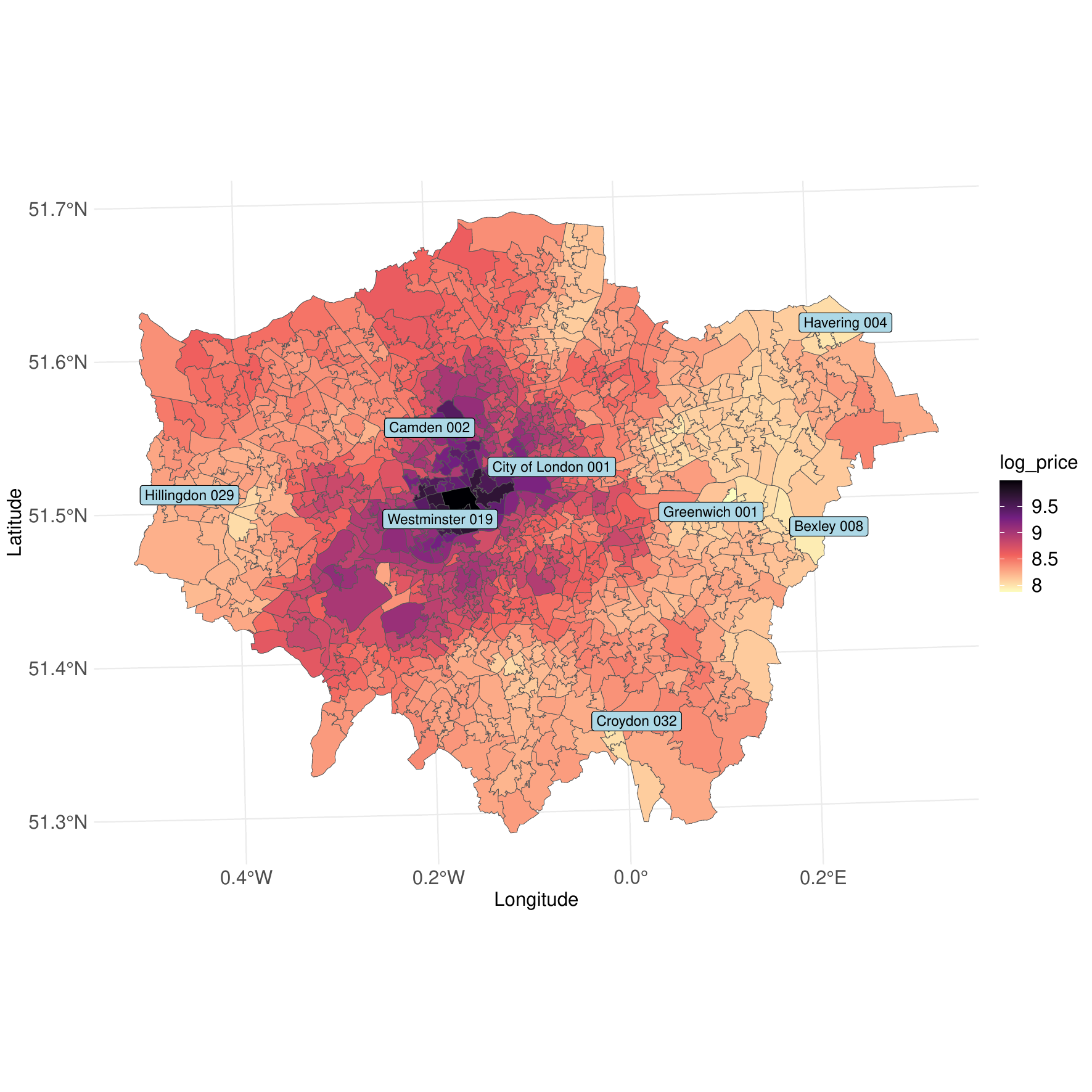}
    \vspace{-2cm}\captionsetup{justification=centering} 
    \caption{log median price per square meter across all time intervals for 983 MSOAs in the London region.}
    \label{fig:median_price}
\end{figure}

Next, the focus is shifted to exploring the spatial relationship in the house prices across the MSOAs in the London region. \Cref{fig:median_price} illustrates the logarithm of the median price per square meter (median taken over the entire observation period) for every MSOA in the dataset. It clearly reveals a spatial trend: central areas like Westminster, Chelsea, Camden and the City of London exhibit higher prices, contrasting with lower prices in the boundaries of the region, e.g., in areas like Bexley, Croydon, Havering and Hillingdon. We find that ``Westminster 019'' has the highest median selling price, while ``Greenwich 001'' records the lowest median selling price in the 106 months of data. Overall, this plot highlights the spatial impact of regions on property prices in London. Empirically, we also observe that the general trend of the median price was increasing until September-October 2017, followed by a decline. 

\begin{figure}[!htb]
    \centering
    \includegraphics[width = 0.6\textwidth,keepaspectratio]{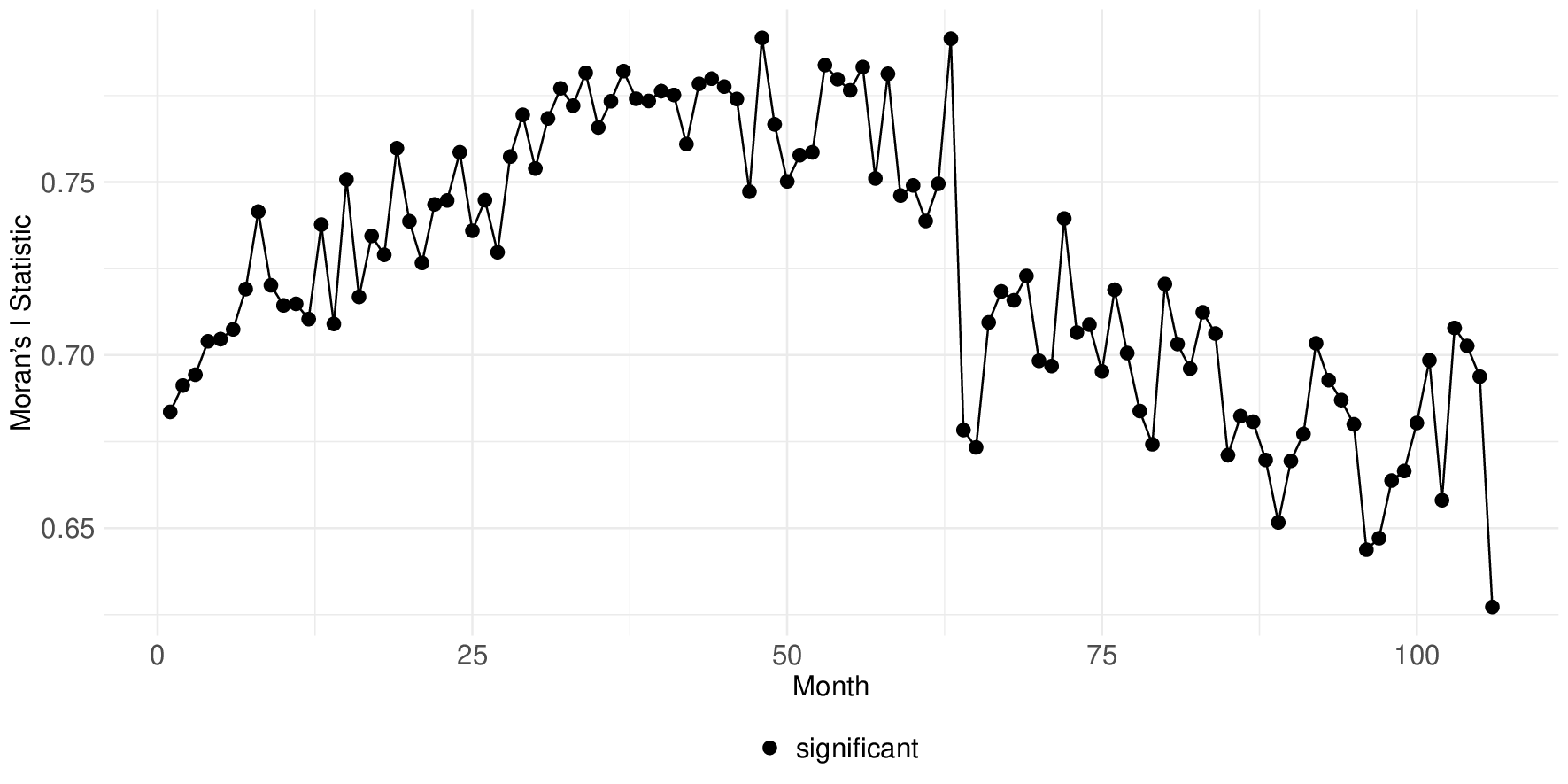}
    %moran_result_plot
    \caption{Moran's I statistic for the log price for all the monthly time intervals in the entire London region. Black circle indicates a significant spatial correlation at 1\% level of significance.}
    \label{fig:moran}
\end{figure}

While the above hints at spatial correlation in the data, we can further assess the significance of the spatial dependence using statistical procedure. To determine this, we employ the Moran's I statistic \citep{moran1950notes} for each month. It is a measure of spatial autocorrelation and can be calculated using %the formula% below, provided that the spatial weight matrix $W$ is correctly defined with zeros on the diagonal:
%\begin{equation}
%\label{eqn : moran}
%I = \frac{S}{S_W}\frac{\sum_{i = 1}^{S} \sum_{j = 1}^{S} %w_{ij}\left( y(s_i,t) - \bar{y_t}\right)\left( y(s_j,t) - \bar{y_t}\right)}{\sum_{i=1}^{983} \left( y(s_i,t) - \bar{y_t}\right)^2}.
%\end{equation}
\begin{equation}
\label{eqn : moran}
    I = \frac{S}{S_W}\frac{\sum_{i = 1}^{S} \sum_{j = 1}^{S} w_{ij}\left( z_i - \bar{z}\right)\left( z_j - \bar{z}\right)}{\sum_{i=1}^{S} \left( z_i - \bar{z}\right)^2},
\end{equation}
where $z_1,\hdots,z_S$ are observations from $S$ number of spatial units, $\bar{z}$ is the sample mean, and $W$ stands for a suitable spatial weight matrix. The weight matrix features zeros on the diagonal and other elements represented by $w_{ij}$, defined as $\exp(-d_{ij})$, where $d_{ij}$ signifies the great circle distance on an ellipsoid between the locations $s_i$ and $s_j$, calculated using the Vincenty method. The term $S_W$ represents the sum of all the elements of $W$. We calculate the Moran's I statistic for the price data at all monthly intervals in London, and present them in \Cref{fig:moran}. There, a black circle denotes statistically significant spatial correlation at 1\% level, and we get clear evidence of a strong spatial correlation across all time-points. 

%Here, with 983 locations, $\bar{y_t}$ denotes the mean of logarithmic price at time $t$ \sd{Did you imply "mean of log-price"?} \kg{Yes} \kg{For ACF, used median} \sd{To avoid confusion, can you just try with mean for ACF as well?}

\begin{figure}[!htb]
    \centering
    \includegraphics[width = 0.6\textwidth]{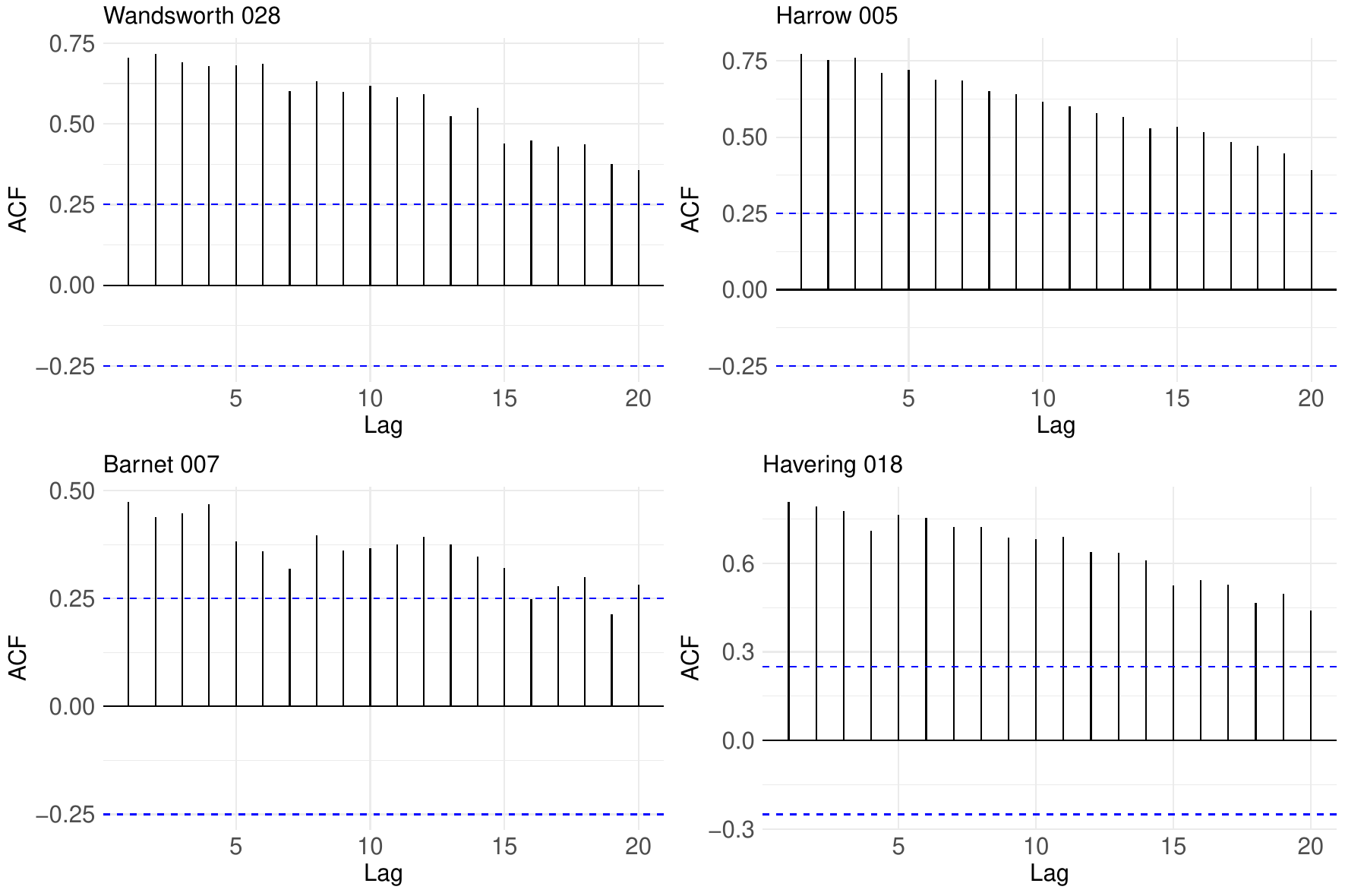}
    \caption{ACF plots for the log price for four randomly selected MSOAs. The dotted lines are corresponding to the critical value at 1\% significance level.}
    \label{fig:ACF}
\end{figure}

Further, to explore the temporal relationships in the price, we compute autocorrelation function (ACF) of the aforementioned variable for individual MSOAs in the London area. Our analysis reveals significant ACF values at multiple lags. As an example, we present the ACF plots for four randomly selected MSOAs in Figure \ref{fig:ACF}. Other regions also depict similar patterns, and the plots are omitted for conciseness. The figure demonstrates a significant temporal correlation in log prices. Considering the pronounced spatial and temporal correlations identified through Moran's I statistics and ACF plots, we proceed to propose a spatio-temporal Gaussian process model for this data.

Note that all computations in this paper are executed on a shared system with 128 GB RAM and 3 GHz 24-core compute nodes, using RStudio Version 2023.12.0.369 equipped with R version 4.3.2. Moran's I statistics is calculated using the ape package \citep{Schliep2019} and the ACF plots are made using the forecast package \citep{hyndman2008automatic}.

%\section{Methods}
%\label{sec:methods}

%\section{Spatio-temporal model for analyzing house-price}

\section{Divide-and-conquer method for spatio-temporal analysis of house price}
\label{sec:sp-model}

Our proposed divide-and-conquer (D\&C) methodology is elucidated in this section. We start by outlining the model suitable to capture the spatio-temporal dynamics of house price, where potentially varying number of transactions are observed at every space-time combination. Subsequently, the entire dataset is split into multiple subsets, and the model is implemented separately for different subsets, bringing to the fore the D\&C framework. These steps, along with the necessary prior specifications and calculations for the Bayesian implementation, are explained in the next two subsections. The last segment of this section provides the details of how to obtain future predictions from the proposed methodology.

%The primary model serves as a method specifically designed for individual partitions. We will then extend this method to different partitions and incorporate additional calculations necessary for the likelihood component. 

\subsection{Main model}
\label{subsec:sp-method}

Throughout this section and after, the term ``location'' represents specific geographical points, denoted by latitude and longitude coordinates $s_i \in \mathbb{R}^2$ (for $i=1,\hdots,S$). As mentioned earlier, in our analysis, these locations correspond to the centroids of the MSOAs. We use $t_j \in \mathbb{N}$ to represent the monthly time-points (for $j=1,\hdots,T$) at which data were recorded. Each combination of location and time-point, $(s_i, t_j)$, is associated with $k_{ij}$ number of observations. The value of $k_{ij}$ is more than 1 when multiple transactions occur, whereas it remains zero when no transaction takes place for the specified location and time point. The latter scenario represents missing data.

We shall use $G = S \times T$ to denote the number of unique space-time combinations in the data, while the total number of observations is given by $N = \sum_{i=1}^{S} \sum_{j=1}^{T} k_{ij}$. Also, wherever used below, $\gauss_k(\cdot, \cdot)$ indicates a $k$-variate Gaussian distribution with suitable mean and variance parameters, $\bm{I}_k$ is the identity matrix of order $k$, and $\mathbb{I}(A)$ is indicator function for set $A$.

%Importantly, $n$ is not simply the product of the total number of locations, $S$, and the total number of time-points, $T$ (i.e., $G = S \times T$). Instead, it significantly surpasses this product.

Let $y(s_i, t_j, r_{ij})$ denotes the logarithm of the price per square meter for the $r_{ij}^{th}$ residential property at location $s_i$ and time-point $t_j$. Our proposed model, for $i = 1, \hdots, S; \; j = 1, \hdots, T; \; r_{ij} = 1, \hdots, k_{ij}$, assumes 
\begin{equation}
    \label{eqn:multiple}
     y(s_i,t_j,r_{ij}) = \mu(s_i,t_j,r_{ij}) + v(s_i,t_j)  +\epsilon(s_i,t_j,r_{ij}), %\quad i = 1, \hdots, S; \; j = 1, \hdots, T; \; r_{ij} = 1, \hdots, k_{ij};
\end{equation}
where $\mu(s_i,t_j,r_{ij})$ denotes the mean structure encompassing an additive combination of covariate effects and $\epsilon(s_i,t_j,r_{ij})$ is an independent and identically distributed white noise process with variance $\sigma_\epsilon^2$. For the spatio-temporal dependence, we assume that all transactions at a given location and time-point share the same characteristics, and thereby consider the same spatio-temporal process for such transactions. This is denoted by $v(s_i,t_j)$, which is assumed to be a zero-mean spatio-temporal process. % For the first term, we propose  the covariate effects. i.e., $v(s_i,t_j,r_{ij}) = v(s_i,t_j)$. $v(s_i,t_j,r_{ij})$ represents  \sd{I'm removing the earlier text about the missingness discussion, it appears to be unclear and I don't think it is needed.} \kg{Okay} %In case of missingness, when we do not observe the $y(s_i, t_j, r_{ij})$, we can still have associated $v(s_i,t_j)$ for that particular location and time combination. Hence, we have to convert the proposed model in matrix form in such a way that it considers the space time process for all the combinations of location and time point, but does not generate any entity for missing/zero observation case.

With these assumptions, the proposed model can be written as
\begin{equation}
    \label{eq:multiple1}
    y(s_i,t_j,r_{ij}) = \beta_0 + \sum_{p = 1}^{m}\beta_p x_p(s_i,t_j,r_{ij}) + v(s_i,t_j)  +\epsilon(s_i,t_j,r_{ij}). %, \quad i = 1, \hdots, S; \quad j = 1, \hdots, T;\quad r_{ij} = 1, \hdots, k_{ij}.
\end{equation}

We can express the above model in matrix form as $\bm{Y} = \mathbb{X}\bm{\beta} + \bm{B}\bm{V} +\bm{\epsilon}$, where $\bm{Y}$ is an $N \times 1$ vector representing all observations of the response variable, arranged first by location and then by time. In a similar fashion, $\mathbb{X}$ is the required $N \times (m+1)$ design matrix with first column having all 1 and subsequent columns having the information of the regressors. The parameter vector $\bm{\beta} = (\beta_0, \beta_1, \hdots, \beta_m)^{\top}$ captures the covariate effects. The term $\bm{B}$ is a sparse $N \times G$ binary matrix wherein every row contains 1 in the column that corresponds to the suitable location-time combination. It is clear that the count of 1s in each column of $\bm{B}$ matches $k_{ij}$ for that specific location-time combination. The count of 1s will be zero in case of no observation for particular location and time point. Finally, $\bm{\epsilon}$ is the $N$-dimensional white noise vector, while the vector $\bm{V}$, with dimensions $G \times 1$, represents the spatio-temporal process. It is formed by concatenating columns extracted from the following matrix:
\begin{equation}
\label{eqn : v_matrix}
    \mathbb{V}_{T\times S} = \begin{bmatrix} v_{s_1,1} & v_{s_2,1} & \dots & v_{s_S,1} \\ v_{s_1,2} & v_{s_2,2} & \dots & v_{s_S,2} \\ \vdots & \vdots & \ddots & \vdots \\ v_{s_1,T} & v_{s_2,T} & \dots & v_{s_S,T} \end{bmatrix}.
\end{equation} 

%where the term $v_{s_i,j}, , i = 1, \hdots, S; , j = 1, \hdots, T;$ denotes the spatio-temporal process associated with location $i$ at time point $j$.

%By using this model, we can focus on individual-level characteristics rather than aggregating all the data. The covariance structure assumes a distance matrix, assuming all properties within a given MSOA have the same location, specifically the centroid of the MSOA. As a result, the number of observations significantly exceeds the distinct combinations of locations and time-points, with $n$ being much larger than $ST$.
In order to account for the spatial-temporal dependence in observed outcomes, $v(s_i, t_j)$ is modeled as a zero-mean Gaussian process. We assume $\bm{V} \sim \gauss_{G}(0,\Sigma)$, where $\Sigma$ is the covariance matrix with a separable structure for the spatial dependence and the temporal dependence. In other words, it is written as a product of purely spatial and purely temporal covariance functions:
\begin{equation}
\label{eqn:cov}
    Cov\left\{ v(s_i,t_u) , v(s_{j},t_{v}) \right\} = \sigma_v^2 \rho_s \left( \norm{s_i - s_{j}}; \phi_s \right) \rho_t \left(\norm{t_u- t_{v}}; \phi_t\right).
\end{equation} 

The separable covariance structure is favored as it significantly reduces the parameters in the covariance matrix, and is a popular choice in similar spatio-temporal studies \citep[see, e.g.,][]{sahu2010fusing, deb2019spatio}. For both spatial and temporal covariance functions, we consider the Mat\'ern structure with the specific choice of $\nu = 0.5$, resulting in the exponentially decaying pattern. Thus, we have:
\begin{equation}
\label{eqn:cov_new}
    \rho_s \left(\norm{s_i - s_{j}};\phi_s \right) = \exp\left(-\phi_s \norm{s_i - s_{j}}\right); \quad
    \rho_t \left(\norm{t_u - t_{v}};\phi_t \right) = \exp\left(-\phi_t \norm{t_u - t_{v}}\right),
\end{equation}
where $\norm{t_u - t_{v}}$ represents the absolute time difference, and $\norm{s_i - s_j}$ is calculated via the Vincenty Ellipsoid method, which computes the great circle distance on an ellipsoid. The latter is chosen for its superior accuracy compared to other approaches \citep{hijmans2021introduction}. It is critical to note that, because of the separability assumption imposed on the joint covariance function, we can write the covariance matrix $\Sigma = \sigma_v^2 \left( \Sigma_s \otimes \Sigma_t\right)$, where $\otimes$ denotes the Kronecker product, and $\Sigma_s$, $\Sigma_t$ respectively denote the spatial and temporal correlation matrices, with %and for $i = 1, \hdots, S$, $j =  1, \hdots, T$, we have 
\begin{equation}
    \label{eqn:sigma_matrices}
    \left(\Sigma_s\right)_{ij} = \exp\left(-\phi_s \norm{s_i - s_{j}}\right); \quad 
    \left(\Sigma_t\right)_{uv} = \exp\left(-\phi_t \norm{t_u - t_{v}}\right).
\end{equation}

\subsection{Prior Specifications}
\label{subsec:Posterior distribution_full}

Let $\bm{\theta} = (\bm{\beta}^{\top}, \sigma_v^2, \sigma_\epsilon^2, \phi_s, \phi_t)^{\top}$ be the vector of unknown parameters. We estimate them completely within a Bayesian framework, primarily using Gibbs sampling \citep{gelfand2000gibbs}, for which suitable prior specifications are essential. For all coefficients in $\bm{\beta}$, we consider independent Gaussian priors with mean 0 and variance $c$, where $c$ is relatively large, typically on the order of $10^4$. For the error variances $\sigma_\epsilon^2$ and $\sigma_v^2$, we employ independent inverse gamma priors $IG(a, \lambda)$, where $a$ and $\lambda$ are the shape and scale parameters, respectively. We choose $a = 2$, making these priors non-informative. We also noticed that the parameter estimates and model selection criteria are insensitive to the initial values of the scale parameter, and throughout this study, we take $\lambda=1$.

For estimating the decay parameters $\bm{\phi} = (\phi_s, \phi_t)$ in the covariance functions, many prior studies \citep[e.g.,][]{sahu2006spatio, rawat2023spatio} have used cross-validation to find the optimal choices. Albeit this approach is computationally more tractable, it has some limitations, as it explores a small set of values and can miss viable solutions. An alternative is the single-variable slice sampling algorithm \citep{neal2003slice}, which provides direct estimation of the parameters within a Bayesian framework. While this method thoroughly explores the parameter space, it demands extensive computation as the covariance matrix must be recalculated and inverted in every iteration of the Bayesian computation. Keeping the advantages of both approaches in view, we aim to achieve a middle ground -- somewhat like a discrete slice sampling -- wherein a discretized sample space is considered for the parameters $(\phi_s,\phi_t)$, and the parameters are estimated directly within the Bayesian setting. This approach allows us to provide credible intervals for these parameters. In terms of the prior specifications, we adopt a discretized uniform prior for $\phi_s$, selecting values from 1 to 3 in intervals of 0.2, resulting in the set $\mathcal{A}_s$. For $\phi_t$, similarly, we select equidistant points between 0.2 and 1 in intervals of 0.2, and denote it as $\mathcal{A}_t$. Below, the product $\mathcal{A}_s \times \mathcal{A}_t$ is indicated as $\mathcal{A}_{st}$, while $\U$ indicates the mentioned uniform distribution.

%This results in 45 combinations derived from $\mathcal{A}_s$ and $\mathcal{A}_t$, denoted as $\mathcal{A}_{st}$. 

%Instead of fitting methods separately as in cross-validation, we calculate the probabilities for each combination of discrete values of $\phi_s$ and $\phi_t$ in each iteration and generate the combination value based on these probabilities. Our discrete setup, termed ``'', pre-calculates these covariance matrices, resulting in improved computational efficiency.

%However, this method has limitations, exploring only a discrete set of values and potentially missing viable solutions. To circumvent that, we adopt the single-variable slice sampling algorithm \citep{neal2003slice} and estimate $\bm{\phi}$ directly within the Bayesian setup. This algorithm allows comprehensive exploration of the parameter space encompassing a wide range of values and avoiding potential omissions. For the decay parameter, we choose uniform prior in the interval $(0,3)$. %\sd{Add a sentence on why these intervals are taken for the priors}. %These priors yield full conditional posterior distributions for the decay parameters during estimation.

As mentioned before, we use concepts of Gibbs sampling and discrete slice sampling to implement the proposed model for estimation and prediction purposes. As the first step, following \Cref{subsec:sp-method}, we can write the full model as
\begin{equation}
    \label{eqn:Y_distribution}
    \bm{Y} \mid \bm{\theta},\bm{V} \sim \gauss_N \left(\mathbb{X}\bm{\beta} + \bm{B}\bm{V}, \sigma_\epsilon^2 \bm{I}_N\right),
\end{equation}
while the prior distributions for all parameters are specified as:
\begin{equation}
\label{eqn:prior}
  \begin{split}
      &\bm{\beta} \sim \gauss_{m+1} \left(0, c\bm{I}_{m+1}\right), \\
      &\sigma_\epsilon^2 \sim IG(a,\lambda), \; \sigma_v^2 \sim IG(a,\lambda), \\
      & (\phi_s, \phi_t) \sim \U(\mathcal{A}_{st}), \\
      &\bm{V} \sim \gauss_G \left(0,\sigma_v^2 \Sigma _s \otimes \Sigma_t\right).
  \end{split}
\end{equation}

Using $f(\cdot)$ to denote the density function in a generic way, the complete posterior distribution is% can now be calculated as 
\begin{equation}
    \label{eqn : joint_post}
f\left(\bm{\beta},\bm{V},\sigma_\epsilon^2,\sigma_v^2,\phi_s,\phi_t \mid \bm{Y}\right) \propto f\left(\bm{Y} \mid \bm{\beta},\bm{V},\sigma_\epsilon^2,\sigma_v^2,\phi_s,\phi_t\right) f\left(\bm{V} \mid \sigma_v^2,\phi_s,\phi_t\right) f(\sigma_v^2)f(\phi_s)f(\phi_t)f(\sigma_\epsilon^2)f(\bm{\beta}).
\end{equation}

For brevity, detailed calculations for the conditional posteriors in case of the complete data are deferred to Appendix A.1 of the supplementary material. In the next subsection, we present the posterior distributions and calculations for the D\&C setup.

\subsection{Divide-and-conquer approach for large dataset}
\label{sec:D&C}

The above model and its Bayesian implementation are helpful in analyzing house price dynamics. However, even with the assumption of separability and the use of Kronecker product to reduce the cost of inverting dispersion matrices, this approach becomes computationally challenging for a dataset with large number of locations or time-points. Specifically, each iteration in the above implementation requires $O(S^2T^2)$ memory units for storage and $O((S \vee T)^3)$ floating-point operations for computation. Naturally, this approach is infeasible for large-scale data, which motivates us to develop a divide-and-conquer approach in this section building upon the above methodology. Although the use of divide-and-conquer is not a new concept in computer science, its implementation in scalable Bayesian inference is a recent advancement. We present the technique in a general three-step distributed framework applicable for scaling posterior computations in our spatio-temporal model for house prices presented in \Cref{subsec:sp-method}.

\subsubsection{First step: partitioning the data}
\label{sec : first_step}

Consider a large dataset with $S$ spatial locations, $T$ time-points, and multiple observations as set out before. Our approach involves splitting these data points into $Q$ non-overlapping subsets for model feasibility. We use a random allocation scheme for the $S$ locations into these $Q$ subsets. This process ensures that each subset contains an adequate number of locations and their corresponding data for all time-points. To avoid bias from locations with significantly more or fewer observations, one should take precautions during the random assignment, striving for balanced representation across the subsets. Our specific approach for the main analysis is elaborated in \Cref{sec:model_matrices}.

% Let $i = 1, 2, \hdots, S$ be the spatial locations, and $j = 1, \hdots, T$ be the time-points, where data is available for each spatial location at each time-point. For each subset $q$ $(q = 1, \hdots, Q)$, we consider a set of $m_q$ locations as $\bm{S}_q = (s_{q1}, s_{q2}, \hdots s_{qm_q})$. Note that a spatial location can be part of multiple subsets, resulting in more partitions. However, in this work, we only consider disjoint subsets. Thus, we have $\sum_{q = 1} ^ {Q} m_q = S$, and the union of the location subsets is equal to the set of all locations. Here, all the time-points are included for each location, so we do not mention the time-points in the detail.

To set the notations for this section, let $\bm{S}_q = (s_{q_1}, s_{q_2}, \hdots s_{q_{m_q}})$ denote the set of $m_q$ locations in the $q^{th}$ subset, $q \in \{1, \hdots, Q\}$. Although in principle a spatial location can be included in multiple subsets, in this work we only consider disjoint partition of the spatial index set, thereby implying $\sum_{q = 1} ^ {Q} m_q = S$ and $\cup_{q=1}^Q \bm{S}_q$ being equal to the set of all locations. In the $q^{th}$ subset, we have the data $\{\bm{Y}_{(q)},\mathbb{X}_{(q)}\}$, with the total number of observations being $N_{(q)} = \sum_{i = 1}^{m_q}\sum_{j = 1}^{T}k_{ij}$. Clearly, $\bm{Y}_{(q)}$ is a $N_{(q)} \times 1$ vector, and $\mathbb{X}_{(q)}$ is a $N_{(q)} \times (m+1)$ matrix representing the covariates corresponding to the observations for all time-points in the spatial locations within the subset $\bm{S}_q$. Following earlier notations, the spatio-temporal Gaussian process model for the $q^{th}$ subset is:
\begin{equation}
\label{eqn:matrix_form_subset}
\bm{Y}_{(q)} = \mathbb{X}_{(q)}\bm{\beta}_{(q)} + \bm{B}_{(q)}\bm{V}_{(q)}+ \bm{\epsilon}_{(q)}.
\end{equation}

% defined as $\bm{Y}_{(q)} = (y_{S_q}(1,t,r_{1t}),\hdots, y_{S_q}(m_q,t,r_{m_q t}))^{\top}$. Similarly, $\mathbb{X}_{(q)}$ is a $N_{(q)} \times (m+1)$ matrix that represents the covariates corresponding to the observation $r_{it}$ ($r_{it} = 1, \hdots, k_{ij}$) in the spatial locations within the subset $\bm{S}_q$ for each time-point $t$ ($t= 1, \hdots, T$). The spatio-temporal Gaussian process model for subset $q$ can be expressed as:
% \begin{equation}
% \label{eqn:subset_model}
% y(s_{qi},t_j,r_{ij}) = \mu(s_{qi},t_j,r_{ij}) + v(s_{qi},t_j) + \epsilon(s_{qi},t_j,r_{ij}), \quad i = 1,\hdots,m_q; \quad j = 1,\hdots,T;\quad r_{ij} = 1, \hdots, k_{ij}.
% \end{equation}

% In matrix form, this can be written as

Throughout the discussions in this section, the parameters $\bm{\theta}_{(q)} = (\bm{\beta}_{(q)}, \sigma_{v(q)}^2, \sigma_{\epsilon(q)}^2, \phi_{s(q)}, \phi_{t(q)})^{\top}$, the data vector $\bm{Y}_{(q)}$, the spatio-temporal process $\bm{V}_{(q)}$, the matrices $\mathbb{X}_{(q)}$, $\bm{B}_{(q)}$, and all the available information $\mathcal{H}_{(q)}$ for subset $q$ are equivalent to their full dataset counterparts $\bm{\theta} = (\bm{\beta}, \sigma_v^2, \sigma_\epsilon^2, \phi_{s}, \phi_{t})^{\top}$, $\bm{Y}$, $\bm{V}$, $\mathbb{X}$, $\bm{B}$, and $\mathcal{H}$. Also, we use the notation $p_q = N/N_{(q)}$ to simplify some expressions. 

\subsubsection{Second step: modified sampling from subset pseudo posterior distributions}
\label{subsec:step2}

We note that the $q^{th}$ subset described above contains a fraction ($1/p_q$) of the full data. Thus, using its posterior will result in an overestimation of the posterior uncertainty compared to using the full dataset. To address this mismatch, we consider a modification to the Bayesian algorithm on the subsets without sacrificing its efficiency. Particularly, a powered likelihood is applied to modify the likelihood of all the parameters before employing the sampling algorithm on the $q^{th}$ subset. As it contains $1/p_q$ fraction of the full data, the asymptotic variances of the posterior distributions of $\bm{\theta}_{(q)}$ and $\bm{V}_{(q)}$ are inflated by a factor of $p_q$ compared to that of $\bm{\theta}$ and $\bm{V}$ \citep{shyamalkumar2022algorithm}. Subsequently, we employ stochastic approximation by raising the likelihood of the parameters $\bm{\beta}_{(q)}, \sigma_{v(q)}^2, \sigma_{\epsilon(q)}^2, \phi_{s(q)}, \phi_{t(q)}, \bm{V}_{(q)}$ in the $q^{th}$ subset to the power of $p_q$ \citep{minsker2014scalable}. These adjustments account for the data present in the other subsets, and ensure accurate estimation of posterior uncertainty while maintaining computational efficiency. 

%Consequently, the likelihoods of the parameters $\bm{\beta}_{(q)}, \sigma_{v(q)}^2, \sigma_{\epsilon(q)}^2, \phi_{s(q)}, \phi_{t(q)}, \bm{V}_{(q)}$ in the $q^{th}$ subset are raised to the power of $N/N_{(q)}$. These adjustments account for the data present in the other $(Q-1)$ subsets.

%However, this inflation factor differs for $\bm{V}_{(q)}$ and $\sigma_{v(q)}^2$ since they involve $ST$ distinct combinations of locations and time-points rather than $n$ in their calculation. Specifically, their asymptotic variance of the posterior distribution is inflated by a factor of $S/m_{q}$ compared to that of $\bm{V}$ and $\sigma_{v}^2$.

  %Following \eqref{eqn:post}, we compute the conditional distribution of the parameters $\delta_{(q)}$ in $\bm{\theta_{(q)}}$ and $\bm{V}_{(q)}$  using pseudo-likelihood as follows:
%\begin{equation}
%\label{eqn:post_pseudo_beta}
%f(\delta_{(q)}|\beta_{(q)}, %\bm{V}_{(q)},\sigma_{\epsilon(q)}^2,\sigma_{v(q)}^2,\phi_{s(q)}, %\phi_{t(q)},\bm{Y}_{(q)}) \propto f(\bm{Y}_{(q)}|\beta_{(q)}, %\bm{V}_{(q)},\sigma_{\epsilon(q)}^2,\sigma_{v(q)}^2,\phi_{s(q)}, %\phi_{t(q)})^{N/N_{(q)}} f(\delta_{(q)}).
%\end{equation}

For the $q^{th}$ subset, as there are $N_{(q)}$ data points, the covariance matrices $\Sigma_{s(q)}$ and $\Sigma_{t(q)}$ have dimensions $m_q \times m_q$ and $T \times T$, respectively. Let $G_{(q)} = m_q \times T$, and $\bm{K}_{(q)}$ be a diagonal matrix of order $G_{(q)} \times G_{(q)}$, where the diagonal elements are $k_{ij}$ for different $(s_i,t_j)$ combinations in the $q^{th}$ subset. The prior distributions for all parameters in subset $q$ align with those in \eqref{eqn:prior}, except for $\bm{V}_{(q)}$, which has the following structure:
\begin{equation}
    \label{eqn : v_q_prior}
    \bm{V}_{(q)}  \sim \gauss_{G_{(q)}} \left(0,\sigma_{v(q)}^2 \left(\Sigma_{s(q)} \otimes \Sigma_{t(q)}\right)\right).
\end{equation}

Utilizing the abovementioned stochastic approximation, we now obtain the following subset pseudo posterior distribution. For the sake of brevity, we omit the technical derivations in the main text, and refer to the supplementary material for detailed information. 

\begin{equation}
\label{eqn : all_q_posterior}
\begin{split}
&\bm{\beta}_{(q)}\mid \mathcal{H}_{(q)} \sim \gauss_{m+1} \left( \left( \frac{p_q\mathbb{X}_{(q)}^{\top}\mathbb{X}_{(q)}}{\sigma_{\epsilon(q)}^2} +  \frac{\bm{I}_{m+1}}{c} \right)^{-1} \left( \frac{p_q\mathbb{X}_{(q)}^{\top} \left(\bm{Y}_{(q)} - \bm{B}_{(q)} \bm{V}_{(q)}\right)}{\sigma_{\epsilon(q)}^2} \right)\left(\frac{p_q\mathbb{X}_{(q)}^{\top} \mathbb{X}_{(q)}}{\sigma_{\epsilon(q)}^2} +\frac{\bm{I}_{m+1}}{c} \right)^{-1}\right),\\
& \sigma_{\epsilon(q)}^2\mid \mathcal{H}_{(q)} \sim IG\left(a+\frac{N}{2}, 
    \frac{p_q\left(\norm{\bm{Y}_{(q)}- \mathbb{X}_{(q)}\bm{\beta}_{(q)} - \bm{V}_{(q)}}\right)^2}{2} + \lambda \right),\\
& \sigma_{v(q)}^2\mid \mathcal{H}_{(q)} \sim IG\left(a+\frac{{p_q G_{(q)}}}{2}, 
    \frac{p_q\left(\bm{V}_{(q)}^{\top}\left(\widetilde{\Sigma}_{s(q)}^{-1} \otimes \Sigma_{t(q)}^{-1}\right) \bm{V}_{(q)}\right)}{2} + \lambda \right),\\
 &   f(\phi_{s(q)}, \phi_{t(q)}) \mid \mathcal{H}_{(q)}\propto 
    |\Sigma_{s(q)}|^{-p_qT/2} |\Sigma_{t(q)}|^{-p_qm_q/2} \exp\left(\frac{-p_q\bm{V}_{(q)}^{\top}\left(\widetilde{\Sigma}_{s(q)}^{-1} \otimes \Sigma_{t(q)}^{-1})\bm{V}_{(q)}\right)}{\sigma_{v(q)}^2}\right)  \mathbb{I}\left((\phi_{s(q)}, \phi_{t(q)}) \in \mathcal{A}_{st}\right), \\
 &  
\bm{V}_{(q)}\mid\mathcal{H}_{(q)} \sim \gauss_{G}\left( \left( \frac{\widetilde{\Sigma}_{s(q)}^{-1} \otimes \Sigma_{t(q)}^{-1}}{\sigma_{v(q)}^2} + \frac{p_q\bm{K}_{(q)}}{\sigma_{\epsilon(q)}^2} \right)^{-1} \left( \frac{p_q\bm{B}_{(q)}^{\top}(\bm{Y}_{(q)}-\mathbb{X}_{(q)}\bm{\beta}_{(q)})}{\sigma_{\epsilon(q)}^2}\right) , \left( \frac{\widetilde{\Sigma}_{s(q)}^{-1} \otimes \Sigma_{t(q)}^{-1}}{\sigma_{v(q)}^2} + \frac{p_q\bm{K}_{(q)}}{\sigma_{\epsilon(q)}^2} \right)^{-1}\right).
\end{split}
\end{equation}

%Here, $d_{\text{max}{(q)}}$ represents the largest distance between two MSOAs in the subset $q$, while $h{(q)}$ denotes the difference between two consecutive values of $\phi_{s(q)}$. 
%, a factor overlooked in previous works such as \cite{guhaniyogi2022distributed} and \cite{guhaniyogi2018meta}. These studies directly considered the inverse of the subset covariance matrix without accounting for spatial interdependencies among subsets. 
%\sd{I think we should avoid making such strong statements, maybe it's good enough to say that we are doing it but not say that others did not do it.} \kg{agree with this, changed the text}

To maintain consistency in our analysis, we adopt identical potential values for decay parameters across all subsets. Therefore, we utilize the same combination of $\phi_{s(q)}$ and  $\phi_{t(q)}$. An essential modification in our posterior calculation involves capturing spatial dependence between subsets. This is achieved by adjusting the inverse of the subset covariance matrix to account for spatial interdependencies among the subsets. To properly follow the procedure, one typically needs to invert the entire spatiotemporal matrix (or the spatial dependence matrix under the separability assumption) before subsetting it for the $q^{th}$ subset. However, inverting such a large matrix is very time-consuming. To address this, we use the concept of Schur's complement \citep{zhang2006schur} and find the inverses of the diagonal matrices efficiently. To elaborate this approach, note that we first obtain the inverse of the last diagonal part using Schur's complement, which involves the inverse of a small partitioned matrix. The other diagonal part is also the inverse of a matrix comprising only the inverse of a small matrix. In the next iteration, we treat this inverse part as the inverse of the updated main matrix and repeat the procedure until only one diagonal matrix remains. By employing this iterative method, we can find the inverses of all the block diagonal matrices without needing to invert the entire matrix. Our extension of the Schur's complement approach significantly speeds up the computation of the inverse. Let us denote the inverse of $\Sigma_{s(q)}$ as ${\widetilde{\Sigma}}_{s(q)}^{-1}$, reflecting the modified version after incorporating spatial dependence. Detailed calculations of this procedure are shown in Appendix B of the supplementary material. 

In our Gibbs sampling procedure, computing the posterior of the vector $\bm{V}_{(q)}$ requires inverting a $G_{(q)} \times G_{(q)}$ matrix in each iteration, which can be computationally demanding. To mitigate this, we adopt a more efficient approach based on the theory of multivariate normal distribution. Specifically, we partition the vector $\bm{V}_{(q)}$ into $m_q$ subvectors based on the columns in the matrix $\mathbb{V}_{(q)}$. Then, we take $\bm{Z^{'}}_1$ as the first subvector of $\mathbb{V}_{(q)}$, and $\bm{Z^{'}}_2$ as a concatenated subvector of the remaining columns of $\mathbb{V}_{(q)}$, i.e., $\bm{V}_{(q)}=\begin{bmatrix} \bm{Z^{'}}_1 \ \bm{Z^{'}}_2 \end{bmatrix}$. Similarly, the covariance matrix $\Sigma_{(q)}$ can be written as 
\begin{equation}
    \Sigma_{(q)} = \sigma_{v(q)}^2 \begin{bmatrix} \Sigma_{11(q)}\otimes \Sigma_{t(q)} \hspace{0.2 cm} \Sigma_{12(q)}\otimes \Sigma_{t(q)} \\ \Sigma_{21(q)}\otimes  \Sigma_{t(q)} 
\hspace{0.2 cm}  \Sigma_{22(q)} \otimes  \Sigma_{t(q)}\end{bmatrix}.
\end{equation}

Then, the conditional distribution of $\bm{Z^{'}}_1$ given $\bm{Z^{'}}_2 = z'_2$ is $\gauss_T(\mu_{c(q)}, \sigma_{v(q)}^2 \Sigma_{c(q)})$, where %$\mu_{c}$ and $\Sigma_{c}$ are given as
\begin{equation}
\begin{split}
    \mu_{c(q)} &= \left(\Sigma_{12(q)}\otimes \Sigma_{t(q)}\right) \left(\Sigma_{22(q)}^{-1} \otimes \Sigma_{t(q)}^{-1}\right)z'_2; \\ 
    \Sigma_{c(q)} &= \left(\Sigma_{11(q)}\otimes \Sigma_{t(q)}\right) - \left(\Sigma_{12(q)}\otimes \Sigma_{t(q)}\right) \left(\Sigma_{22(q)}^{-1} \otimes \Sigma_{t(q)} ^{-1}\right) \left(\Sigma_{21(q)}\otimes \Sigma_{t(q)}\right).
\end{split}
\end{equation}

Using the properties of the Kronecker product, the expression can be further simplified as

\begin{equation}
    \mu_{c(q)} = \left(\left(\Sigma_{12(q)} \Sigma_{22(q)}^{-1}\right) \otimes \bm{I}_T\right)z'_2; \quad
    \Sigma_{c(q)} = \left(\Sigma_{11(q)} - \Sigma_{12(q)} \Sigma_{22(q)}^{-1}\Sigma_{21(q)}\right)\otimes  \Sigma_{t(q)}.
\end{equation}

%we divide the vector $\bm{V}_{(q)}$ into $\bm{V}_{1(q)}, \bm{V}_{2(q)}, \hdots, \bm{V}_{m_q(q)}$. The conditional distribution of $\bm{V}_{i(q)}$ consists the terms involving only $\bm{V}_{i(q)}$, then get the following distribution for $\bm{V}_{1(q)}$ as
 
Now, following the above, we can write
\begin{equation}
    \begin{split}
    \bm{V}_{1(q)} \mid\mathcal{H}_{(q)}  & \sim \gauss_T \left( \left( \frac{\Sigma_{c(q)}^{-1}}{\sigma_{v(q)}^2} + \frac{p_q \bm{K}_{1(q)}}{\sigma_{\epsilon(q)}^2} \right)^{-1} \left( \frac{\Sigma_{c(q)}^{-1} \mu_{c(q)}}  {\sigma_{v(q)}^2} + \frac{p_q \left(\bm{B}_{1(q)}^T \left(\bm{Y}_{1(q)}-\mathbb{X}_{1(q)}\bm{\beta}_{(q)}\right)\right)}{\sigma_{\epsilon(q)}^2}\right) \right., \\
    & \hspace{1in} \left.\left( \frac{\Sigma_{c(q)}^{-1}}{\sigma_{v(q)}^2} + \frac{p_q \bm{K}_{1(q)}}{\sigma_{\epsilon(q)}^2} \right)^{-1} \right),
    \end{split}
\end{equation}
where $\bm{Y}_{1(q)}$ be the data for the first location and all time-points from the $q^{th}$ subset. The matrix $\mathbb{X}_{1(q)}$ is defined accordingly. $\bm{K}_{1(q)}$ is a $T \times T$ diagonal matrix with each diagonal element being equal to $k_{1j}$ for each $(s_{q_1}, t_j)$. $\bm{B}_{1(q)}$ is a submatrix of $\bm{B}_{(q)}$, consisting of the first $T$ columns and $\sum_{j=1}^{T} k_{1j}$ rows, resulting in a sparse matrix of size $\sum_{j=1}^{T} k_{1j} \times T$.
Other notations have the same definitions as before.  We can compute the conditional posterior distributions of $\bm{V}_{2(q)}, \hdots, \bm{V}_{m_q(q)}$ in an identical fashion. Finally, to compute the posterior distribution of the full vector $\bm{V}_{(q)}$, we iteratively employ the conditional distributions in a Gibbs sampler. This approach generates independent posterior samples after an initial burn-in period. Consequently, the posterior distribution of $\bm{V}_{(q)}$ is computed efficiently without incurring the computational cost of inverting the $G_{(q)} \times G_{(q)}$ matrix in each Gibbs sampler iteration. It is imperative to point out that these subset pseudo posteriors significantly speed up the computations and can be run in parallel  for all partitions.

\subsubsection{Third step: combination of subset posteriors}
\label{subsec : combine}

The final and most crucial step of the D\&C algorithm is to combine the subset posterior draws. As discussed in \Cref{sec:introduction}, various existing methods can be employed in this regard. We follow the Wasserstein Barycenter (WB) technique to approximate the complete posterior distribution. This approach has been successfully utilized by \cite{li2017simple} and \cite{srivastava2018scalable} in the setting of independent data, whereas \cite{ou2021scalable} demonstrated the accuracy of using this for combining subset posteriors in the case of time series data. Motivated by these studies, we apply a similar rationale for spatio-temporal data and utilize the WB to combine the subset posteriors. It is worth noting that Wasserstein-based posteriors exhibit appealing asymptotic properties, as shown by \cite{szabo2019asymptotic}. 

Before going into the detail, it is critical to discuss the mathematical underpinning of the Wasserstein distance. Let $(M,d)$ be a complete separable metric space, $\mathcal{P}(M)$ be the space of all probability measures on $M$ and $\mathcal{P}_2(M)$ be the set of all probability measures on $M$ with finite second moments. For $\varrho, \xi \in \mathcal{P}_2(M)$, $\Pi(\varrho,\xi)$ denotes the set of all probability measures on $M \times M$ with marginals $\varrho$ and $\xi$. Then, with $d(\cdot,\cdot)$ indicating the euclidean metric, the Wasserstein distance of order 2 between $\varrho$ and $\xi$ is defined as
\begin{equation}
\label{eqn:WB}
   W_2(\varrho,\xi) = \left\{\inf_{\pi \in \Pi(\varrho,\xi)} \int_{M\times M} d(x_1,x_2)^2 \Pi(x_1,x_2)\right\}^{1/2}.
\end{equation}
%\begin{equation}
%   \label{eqn:WB}
%   W_2(\mu,\nu) = \left\{\inf_{\pi \in \Pi(\mu,\nu)} \int_{M\times M} \norm{x_1 - %x_2}^2 \Pi(x_1,x_2)\right\}^{1/2}.
%\end{equation} and $d(x_1,x_2) = \norm{x_1 - x_2}^2$

As \cite{bickel1981some} discussed, the convergence in $W_2$ on $\mathcal{P}_2(M)$ implies  weak convergence of probability measures and convergence of the second moment. Keeping that in view, if $\bm{\theta}_{(1)}, \hdots, \bm{\theta}_{(Q)}$ are the $Q$ subset posterior distributions, then the approximated posterior $\hat{\bm{\theta}}$ is defined as
\begin{equation}
    \label{eqn:Wasp}
    \hat{\bm{\theta}} = \argmin_{\bm{\theta} \in \mathcal{P}_2(M)}\frac{1}{Q}\sum_{i = 1}^{Q}W_2^2(\bm{\theta},\bm{\theta}_{(i)}).
\end{equation}

This approach is referred to as the Wasserstein Posterior (WASP). The exact computation of the WASP is known to be computationally intensive and remains a subject of ongoing research. We adopt an approximation for $\hat{\bm{\theta}}$, following \cite{shyamalkumar2022algorithm}. This completes the last step in our proposed approach (D\&C-STBD), and the pseudocode of the entire procedure is shown in \Cref{alg:EDC-STBD}. 

\begin{algorithm}[!htb]
\caption{Proposed divide-and-conquer methodology for large spatio temporal data (D\&C-STBD)}
\label{alg:EDC-STBD}
 \SetKwInOut{Input}{Input}
    \SetKwInOut{Output}{Output}
\Input{Spatio temporal dataset (Y,$\mathbb{X}$), where Y is the response vector and $\mathbb{X}$ is the complete information on regressors. Number of subsets $Q$, the total number of post burn-in iterations on every subset is $L$, size of each subset, prior distribution for all the parameters.}
\Output{
\begin{enumerate}
    \item $\hat{\bm{\theta}}^{(1)}, \ldots , \hat{\bm{\theta}}^{(QL)}$ as approximate WASP draws.
    \item The approximated $\gamma$ quantile for $\bm{\theta}$ is then calculated as: 
    \begin{equation*}
        \label{eqn:quantile_algo}
        \hat{\bm{\theta}}^\gamma = \frac{1}{QL}\sum_{i = 1}^{QL} \hat{\bm{\theta}}^{(i)\gamma}.
    \end{equation*}
\end{enumerate}
}
 {\bf Algorithm:}
\begin{enumerate}
    \item Divide data into $Q$ subsets, with $q^{th}$ subset as $(\bm{Y}_{(q)},\mathbb{X}_{(q)})$, each containing data from different locations but all time-points.
    \item Compute the subset posteriors for all parameters of all subsets as discussed in \Cref{subsec:step2}. 
    \item \textbf{for} $q = 1,\hdots Q$ do\\
         Obtain subset posterior draws (every tenth MCMC sample after convergence) for all parameters of all subsets at $l^{th}$ iteration using Gibbs and slice sampling: $\bm{\theta}_{(q)}^{(l)} (l = 1,\hdots, L )$.\\
    \textbf{end for}
    \item Compute mean vectors and covariance matrices of subset posterior distributions and the approximate WASP posterior distribution: $\hat{\mu}_{(q)}, \hat{\bar{\mu}}, \hat{\Sigma}_{(q)}, \hat{\bar{\Sigma}}$.
    \item Center and scale subset posterior draws for $q = 1, \hdots, Q; l = 1,\hdots, L $:
    \begin{equation*}
        \hat{c}_{(q)}^{(l)} = \hat{\Sigma}^{-1/2}_{(q)} \left(\bm{\theta}_{(q)}^{(l)} - \hat{\mu}_{(q)}\right)
    \end{equation*}
    \item Define combined posterior using $q^{th}$ subset posterior draws of $\bm{\theta}$:
    \begin{equation*}
        \hat{\bm{\theta}}^{(l^{'})} =  \hat{\bar{\mu}} + {\hat{\bar{\Sigma}}}^{1/2} \hat{c}_{(q)}^{(l)}, \hspace{0.1 cm} l^{'} = (q-1)L + l,
    \end{equation*}
    where $\hat{\bm{\theta}}^{(l^{'})}$ is the approximate WASP draw for $\bm{\theta}$.
\end{enumerate}
\end{algorithm}

A brief outline of the WASP procedure is also of the essence here. First, posterior draws of $\bm{\theta}_{(1)}, \hdots, \bm{\theta}_{(Q)}$ are taken from the $Q$ subsets using the Bayesian algorithm developed earlier. In all instances of Gibbs and discrete slice samplers, the convergence of the chains are confirmed through the diagnostic method of \cite{geweke1991evaluating}. After convergence, we apply thinning to collect posterior samples for each subset. These sample points (say, $L$ for each subset) are then consolidated into a unified set of $QL$ samples. Using these posterior samples, we compute mean vectors $\hat{\mu}_{(q)}$ and covariance matrices $\hat{\Sigma}_{(q)}$ for each subset. This results in a total of $Q$ mean vectors and covariance matrices.  Next, we take the mean of these and get the overall mean $\hat{\bar{\mu}}$ and covariance matrix $\hat{\bar{\Sigma}}$. Now, using the subset mean vectors and covariance matrices, we standardize and center the subset posterior samples. Then, we use the overall mean vector and covariance matrix to get the approximate WASP draws for the actual posterior draws, which leads to $QL$ number of WASP posterior samples for each parameter. We estimate both the posterior mean of these parameters and their corresponding credible intervals using the consolidated sample. On the other hand, for the final estimate of $\bm{V}$ from the subset posterior samples of $v_{(q)}$, we calculate the mean of the samples within each subset to get an estimate for that subset. This leads to $Q$ number of posterior vectors, each corresponding to a different location and time-point. Finally, these vectors are stacked to form the final estimate of $\bm{V}$.
%For instance, the $q^{th}$ posterior vector obtained in this manner is $(\hat{v}_{s_{q_1},1}, \hdots, \hat{v}_{s_{q_{m_q}},T})$ \sd{This notation is very hard to understand, I think we better remove this entire sentence}. 

%\begin{equation}
%    \label{eqn:V_combine_post}
%   \hat{\mathbb{V}} =  \begin{bmatrix} \hat{v}_{s_{11},1} %&  \hat{v}_{s_{21},1} & \dots & \hat{v}_{s_{Q1},1} \\ 
%   \hat{v}_{s_{11},2} &  \hat{v}_{s_{21},2} & \dots & %\hat{v}_{s_{Q1},2} \\ 
%   \vdots & \vdots & \ddots & \vdots \\ %\hat{v}_{s_{1m_1},T} &  \hat{v}_{s_{2m_2},T} & \dots & %\hat{v}_{s_{Qm_Q},T} \end{bmatrix}
%\end{equation}

%The above matrix is for demonstration purposes only and may not form a valid matrix when the lengths of the vectors $m_q, q=1,\hdots, Q$ are not equal. We use the the 2.5\%, 50\%, and 97.5\% quantile sample value for each $\theta_q, q = 1,\hdots, k$, and the \eqref{eqn:quantile} to obtain the posterior median and 95\% credible interval. 

%\begin{equation}
%    \label{eq:phi_s}
%    \phi_s|\bm{V},\sigma_v^2,\phi_t \propto |\Sigma_s|^{-T/2} \exp\left(-%\frac{\bm{V}^{\top}\left( \Sigma_s^{-1} \otimes \Sigma_t^{-1}\right) V}%{2\sigma_v^2}\right) 
%\end{equation}
%
%\begin{equation}
%    \label{eq:phi_t}
%    \phi_t|\bm{V},\sigma_v^2,\phi_s \propto |\Sigma_t|^{-S/2} \exp\left(-%\frac{\bm{V}^{\top}\left( \Sigma_s^{-1} \otimes \Sigma_t^{-1}\right) V}%{2\sigma_v^2}\right) 
%\end{equation}

\subsection{Prediction}
\label{subsec : pred}

Our defined spatio-temporal Gaussian process model offers the ability to predict the outcome variable for any spatial location and time-point, which are not observed in the data. This is the primary advantage of this method. Specifically, for a residential property $r'$ in a new location $s'$ at a future time-point $t'$, the predicted outcome variable $y(s',t', r')$ is conditionally independent of the observed outcomes $\bm{Y}$ given the spatio-temporal Gaussian process model $v(s',t')$ and the estimated $\bm{\beta}$. If $\bm{X}(s',t',r')$ is the covariate vector for the new observation, then the predicted outcome variable follows a Gaussian distribution: %with mean $\mu(s',t',r') + v(s',t')$ and variance $\sigma_\epsilon^2$.
\begin{equation}
\label{eqn:prediction}
    y(s',t',r') \mid v(s',t') \sim \gauss\left(\bm{X}^\top (s',t',r') \hat{\bm{\beta}} + v(s',t'), \sigma_\epsilon^2\right),
\end{equation}
%where $\bm{X}(s',t',r')$ is the vector of order $(p+1)$  with the first entry being 1 and the subsequent entries containing the regressors' information for the combination $(s',t',r')$.

We estimate the model parameters using a Gibbs sampler on the training data, yielding posterior estimates for $\bm{\beta}$ and $\sigma_\epsilon^2$. To find the conditional distribution of $v(s',t')$ given $\bm{V}$, we use the posterior value of $\bm{V}$ from training data estimates and apply the multivariate Gaussian theory related to the conditional distribution. We treat $v(s',t')$ as the first subvector of the multivariate joint distribution, and $\bm{V}$ obtained from the training dataset as the second subvector. The joint distribution of $v(s',t')$ and $\bm{V}$ is given by
\begin{equation}
    \begin{pmatrix}
        v(s',t')\\
        \bm{V}
    \end{pmatrix} \sim \gauss_{G+1}\begin{pmatrix}
        \begin{pmatrix}
        0\\
        0
    \end{pmatrix},\sigma_v^2 \begin{bmatrix} 1 & \Sigma'^\top   \\ \Sigma'  & \Sigma_s \otimes \Sigma_t\end{bmatrix}
    \end{pmatrix}, \; \text{with} \; \Sigma' = \Sigma_s(s-s') \otimes \Sigma_t(t-t'),
\end{equation}
where $\Sigma_{s(s-s')}$ is a vector of order $S$ with $i^{th}$ entry as $\rho_s \left( ||s_i - s'||;\phi_s \right)$ and $\Sigma_{t(t-t')}$is a vector of order $T$ with $j^{th}$ entry as $\rho_t \left( ||t_j - t'||;\phi_t \right)$. The conditional distribution is then given by
\begin{equation}
    \label{eqn:cond_dis1}
    v(s',t') \mid \bm{V} \sim \gauss\left(\Sigma' (\Sigma_s^{-1} \otimes \Sigma_t^{-1})\bm{V},
    \sigma_v^2 (1 - \Sigma' (\Sigma_s^{-1} \otimes \Sigma_t^{-1})\Sigma'^\top)\right).
\end{equation}
%also normal and follows $N(\mu^', \sigma_v^2 \Sigma')$, where $\mu'$ and $\Sigma'$ are %given as
%\begin{equation*}
%\mu' = \Sigma' (\Sigma_s^{-1} \otimes \Sigma_t^{-1})\bm{V}
%\end{equation*}
%\begin{equation*}
%\Sigma' = 1 - \Sigma' (\Sigma_s^{-1} \otimes \Sigma_t^{-1})\Sigma'^\top
%\end{equation*}

Thus, by predicting $v(s',t')$ using the above conditional distribution and the estimates of $\bm{\beta}$ from the Gibbs sampling procedure, we can predict $y(s',t',r')$ using \eqref{eqn:prediction}.

Now, we shift our focus to the prediction stage in D\&C case. Keeping parity with the earlier notations, we need to forecast $v(s',t')$ for a location $s'$ and a time-point $t'$ in the test dataset. Our training dataset is divided into $Q$ subsets, each treated as a separate training dataset. For each subset, we predict $v(s',t')$ using a similar process as for the whole dataset, resulting in $Q$ realizations of $v(s',t')$. These realizations are denoted as ${v}_q(s',t')$, for $1 \leqslant q \leqslant Q$. The joint distribution of ${v}_q(s',t')$ and $\bm{V_{(q)}}$ is given by 
\begin{equation}
    \begin{pmatrix}
        {v}_q(s',t')\\
        \bm{V_{(q)}}
    \end{pmatrix} \sim \gauss_{G_{(q)}+1}\begin{pmatrix}
        \begin{pmatrix}
        0\\
        0
    \end{pmatrix}, {\sigma_v}_{(q)}^2 \begin{bmatrix} 1 & \Sigma_{(q)}'^\top   \\ \Sigma_{(q)}'  & \Sigma_{s(q)} \otimes \Sigma_{t(q)}\end{bmatrix}
    \end{pmatrix}, \; \text{with} \; \Sigma_{(q)}' = \Sigma_{s(q)}(s-s') \otimes \Sigma_{t(q)}(t-t'),
\end{equation}
where $\Sigma_{{s(q)}(s-s')}$ is a vector of order $S_{(q)}$ with $i^{th}$ entry as $\rho_s \left( ||s_i - s'||;\phi_{s(q)} \right)$ and $\Sigma_{t(t-t')}$is a vector of order $T$ with $j^{th}$ entry as $\rho_t \left( ||t_j - t'||;\phi_{t(q)} \right)$. The conditional distribution is then given by
\begin{equation}
    \label{eqn:cond_dis2}
    {v}_q(s',t') \mid  \bm{V_{(q)}} \sim \gauss\left(\Sigma_{(q)}' ({\widetilde{\Sigma}}_{s(q)}^{-1} \otimes \Sigma_{t(q)}^{-1})\bm{V_{(q)}},
    \sigma_{v(q)}^2 (1 - \Sigma_{(q)}' ({\widetilde{\Sigma}}_{s(q)}^{-1} \otimes \Sigma_{t(q)}^{-1})\Sigma_{(q)}'^\top)\right).
\end{equation}

With the above predicted values of  ${v_{(q)}}(s',t')$ and the estimated parameters within each subset $q$, we can predict the response variable, i.e.\ the price of a house in logarithmic scale, for any MSOA at a future time-point in the test data using \eqref{eqn:prediction_d&c}, which is based on the $q^{th}$ subset data.
\begin{equation}
\label{eqn:prediction_d&c}
    y_{(q)}(s',t',r') \mid v_{(q)}(s',t') \sim \gauss\left( \bm{X}_{(q)}^\top (s',t',r') \hat{\bm{\beta}}_{(q)} + v_{(q)}(s',t'), {\sigma_\epsilon}_{(q)}^2\right).
\end{equation}

Finally, to predict $v(s',t')$ and the response variable $y(s',t',r')$ using the entire training dataset, we take the median of ${v}_q(s',t')$ and $y_{(q)}(s',t',r')$ respectively, for $1 \leqslant q \leqslant Q$. This approach leverages the D\&C framework to efficiently handle large datasets by breaking them into smaller, manageable subsets while maintaining prediction accuracy. It is noteworthy that we incorporate the original inverse of the spatial covariance matrix for each subset to address spatial dependence.

\section{Model implementation and evaluation}
\label{sec:model_matrices}

In this section, we explore the key implementation steps for our model and discuss other competing models along with the evaluation criteria used for benchmarking. 

Recall that our dataset comprises data on $N=651202$ transactions from $S=983$ locations and $T=106$ time-points. As covariates, we include the variables listed in \Cref{tab:variables}, along with linear and quadratic time trends, as well as the interaction of time with carbon emissions. We take logarithmic transformation for all the numeric variables. In the first step of our methodology, the dataset is partitioned into $Q = 20$ disjoint subsets. The initial 19 subsets are comprised of randomly selected 49 locations, while the last subset contains the remaining 52 locations. Each subset encompasses data for all the time-points and their corresponding multiple records. Next, for each subset, we employ the model specified in \eqref{eqn:matrix_form_subset} and fit the models separately using the posterior calculations shown there. After convergence of the chains, 2000 MCMC samples are taken in each of the $Q$ subsets. Finally, we adopt the subset posterior combination method, as described in \Cref{subsec : combine}, to obtain final estimates for all the parameters. Additionally, we determine the values of $v(s,t)$ corresponding to each combination of $(s,t)$ in the dataset.

% To facilitate parameter estimation, we utilize non-informative priors: $\gauss_{17}(0, 10000\bm{I}_{17})$ for $\bm{\beta}$, $IG(2,1)$ for $\sigma_v^2$ and $\sigma_\epsilon^2$, $U(0,3)$ for $\phi_s$ and $\phi_t$ for the $q^{th}$ subset. To estimate the parameters, we employ Gibbs and slice samplers for each subset, ensuring convergence. We obtain 2000 MCMC samples, taking every tenth sample to ensure independence between observations. This process provides us with posterior samples from each of the $q$ subsets. Finally, we adopt the subset posterior combination method, as described in \Cref{subsec : combine}, to obtain final estimates for all the parameters. Additionally, we determine the values of $v(s,t)$ corresponding to each combination of $(s,t)$ in the dataset.

To assess the fit of the proposed approach, we conduct a comparative analysis with four alternative models. The first model is a standard hedonic regression that employs the same covariates as our approach. As discussed in \Cref{sec:introduction}, it is one of the most popular approaches in related research. The second model incorporates an additional fixed effect for the geographic region where the MSOA is situated. We refer to this model as the spatial hedonic model, and it closely parallels the approach described in \cite{sommervoll2019learning}. While these two approaches are conventional techniques in the extant literature of house price analysis, we include two other models in the comparison study for understanding the effectiveness of the proposed spatial and temporal dependence structures. The third model incorporates temporal dependence in a manner similar to the main model proposed in this paper, but without the spatial component. Given that there are 106 time points, the inversion of a $106 \times 106$ matrix during iteration is computationally efficient, making this model's computation manageable without the D\&C step. The fourth model, on the other hand, does not consider temporal dependence but introduces only a spatial error process. With 983 locations, matrix inversion in this model could be time-consuming in each iteration. Therefore, this spatial error process model is integrated into the proposed D\&C framework and implemented in the proposed fashion. This model will be referred to as D\&C-SBD. By comparing our method with these models, we aim to assess their respective data-fitting capabilities. Specifically, we evaluate the accuracy of each model's fit to the data and examine whether the resulting estimates are consistent with existing literature. This comparison will help determine the robustness and reliability of our approach in relation to established models.

We evaluate the efficacy of these models by fitting the data. Label the entire dataset as $\mathcal{D}$, with an observation for location $s_i$ at time-point $t_k$ be denoted as $y(s_i, t_k, r_{ik})$. The corresponding fitted value is $\hat y(s_i, t_k, r_{ik})$. To understand the goodness of fit of every model, we start with the ubiquitous coefficient of determination ($R^2$) which quantifies the amount of variability in the response variable that can be accounted for by the predictor variables, thereby indicating the degree of adequacy of the model. Further, we consider the mean absolute error (MAE), mean absolute percentage error (MAPE) and root mean square error (RMSE) of the fitted values, as recommended in statistical literature \citep{botchkarev2019new}. The MAE, i.e.\ the average absolute difference between actual and fitted values, provides a sense of the extent of the errors without considering their direction (overestimation or underestimation). MAPE, on the other hand, computes the percentage deviation and offers a straightforward interpretation of the relative error. RMSE is one of the most popular metrics and is advantageous because it penalises large errors more severely than small errors, thereby providing a more impartial evaluation of the models' accuracy. Finally, in some cases, a model with a lower average error (as measured by MAE, MAPE, or RMSE) may exhibit higher variability in errors. To that end, the variance of the absolute error (VAE) can help to assess the robustness or stability of the model's performance, especially in scenarios where consistent predictions are essential.

In general, a model with lower MAE, MAPE, RMSE, and VAE values is regarded as more accurate. Mathematically, these are defined as: %y(s_i, t_k,r_{ik}) \in 
 \begin{equation}
 \label{eqn: matrices}
 \begin{split}
 &\mathrm{MAE} = \frac{1}{\abs{\mathcal{D}}} \sum_{\mathcal{D}} \abs{\hat y(s_i, t_k,r_{ik})- y(s_i, t_k,r_{ik})}, \\
    & \mathrm{MAPE} =\frac{100\%}{\abs{\mathcal{D}}} \sum_{ \mathcal{D}} \abs{\frac{\hat y(s_i, t_k,r_{ik}) -  y(s_i, t_k,r_{ik})}{y(s_i, t_k,r_{ik})}},\\
     &\mathrm{RMSE} = \sqrt{\frac{1}{\abs{\mathcal{D}}} \sum_{\mathcal{D}}(\hat y(s_i, t_k,r_{ik}) -  y(s_i, t_k,r_{ik}))^2},\\
    &\mathrm{VAE} = \frac{1}{\abs{\mathcal{D}}} \sum_{\mathcal{D}} \left(\abs{\hat y(s_i, t_k,r_{ik})- y(s_i, t_k,r_{ik})} - \mathrm{MAE}\right)^2.
 \end{split}
\end{equation}

% that calculates the square root of the average of squared errors to determine the difference between the actual and predicted values. RMSE is advantageous because it penalises large errors more severely than small errors, thereby providing a more impartial evaluation of the model's accuracy. In general, a model with lower MBE, MAE, MAPE, and RMSE values is regarded as more accurate.

% \begin{equation}
%     \label {eqn : r_square}
%     \mathrm{R^2} = 1-\frac{\sum_{i=1}^{N}(y_i - \hat{y_i})^2}{\sum_{i=1}^{N}(y_i - \bar{y_i})^2},
% \end{equation}
% where $y$ is actual log price per square meter value, $\hat{y}$ is the predicted value and $N$ is the size of the dataset.

Finally, we steer to the prediction accuracy of our proposed model. In this case, the whole data $\mathcal{D}$ is split into a training set (call it $\mathcal{D}_{\mathrm{tr}}$) and a test set (call it $\mathcal{D}_{\mathrm{te}}$). We fit the proposed model to $\mathcal{D}_{\mathrm{tr}}$ and use the estimated parameters to predict the log house price per square meter for each entry in $\mathcal{D}_{\mathrm{te}}$. In this case, using the true and predicted values for $\mathcal{D}_{\mathrm{te}}$, we primarily calculate the mean absolute percentage error to understand the forecasting performance of the model. As we illustrate in \Cref{subsec:prediction}, by considering different types of test sets, we can analyze and comment on the robustness of the predictive capability of our methodology.

%Additionally, we also consider the mean bias error (MBE), which provides an idea of how near the predicted values are to the actual values, and is defined as
%\begin{small}
%\begin{equation}
    %\mathrm{MBE} = \frac{1}{|\mathcal{D}_{\mathrm{te}}|} \sum_{y(s_i, t_k,r_{ik}) \in \mathcal{D}_{\mathrm{te}}} \left(\hat y(s_i, t_k,r_{ik})- y(s_i, t_k,r_{ik})\right).
%\end{equation}
%\end{small}
%Evidently, a value close to 0 is desired. A positive MBE for a model indicates the tendency of over-forecasting whereas a negative value has the opposite interpretation.

%\section{Results and discussions}
\section{Results and discussions}
\label{sec:results}

\subsection{Estimated model}

We begin by presenting the estimates of the parameters obtained from fitting our proposed model to the entire data. The results are shown in \Cref{tab:model_result}, which displays the posterior means of the parameters along with their 95\% credible intervals. It helps us carefully examine the key factors influencing property prices in the London region, and provide valuable insights into the complex dynamics underlying property pricing. To begin, we note that the credible intervals for all the covariates used in our study, except for the presence of wind turbine, are away from zero, thereby suggesting significant impact of these factors.

\begin{table}[hbt!]
\centering
\caption{Parameter estimates (posterior mean) and the corresponding credible intervals (CI) when the proposed model is fitted to the whole data with 983 MSOAs and 106 time-points (651202 observations).}
 \renewcommand{\arraystretch}{1.2}
 \label{tab:model_result}
\begin{tabular}{lcc}
  \toprule
 Variable & Estimate & 95\% CI \\ 
  \midrule
Intercept & 9.675 & (9.640 , 9.711) \\ 
  Linear time trend & 0.982 & (0.845 , 1.103) \\ 
  Quadratic time trend & $-0.646$ & $(-0.757 , -0.534 )$ \\ 
  log (1 + Area) & $-0.319$ & $(-0.321 , -0.316)$ \\ 
  Rooms (3 Rooms) & 0.066 & (0.065 , 0.068) \\ 
  Rooms (4 Rooms) & 0.059 & (0.057 , 0.062) \\ 
  Rooms (5 Rooms) & 0.064 & (0.061 , 0.067) \\ 
  Rooms ($>$ 5 Rooms) & 0.109 & (0.106 , 0.113) \\ 
  Property type (Terraced) & $-0.052$ & $(-0.054 , -0.050)$ \\ 
  Property type (Detached) & 0.141 & $(0.138 , 0.144)$ \\ 
  Property type (Flats) & $-0.250$ & $(-0.252 , -0.247)$ \\ 
  Fireplace & 0.064 & $(0.062 , 0.065)$ \\ 
  Ventilation (Extract-only) & 0.029 & (0.022 , 0.036) \\ 
  Ventilation (Both supply and extract) & 0.055 & (0.043 , 0.065) \\ 
  Wind turbine & 0.006 & ($-0.019$ , 0.031) \\ 
  log (1 + CO2 emission) & $-0.031$ & $(-0.035 , -0.028)$ \\ 
  Interaction of trend and CO2 emission & 0.076 & (0.070 , 0.081) \\ 
  $\sigma_\epsilon^2$  & 0.043 & (0.043 , 0.043 )\\ 
  $\sigma_v^2$ & 0.083 & (0.070 , 0.103) \\ 
  $\phi_s$ & 2.402 & (1.486 , 3.041) \\ 
  $\phi_t$ & 0.528 & (0.183 , 0.824) \\
   \bottomrule
\end{tabular}
\end{table}

\begin{figure}[!htb]
    \centering
    \includegraphics[width = 0.68\textwidth,keepaspectratio]{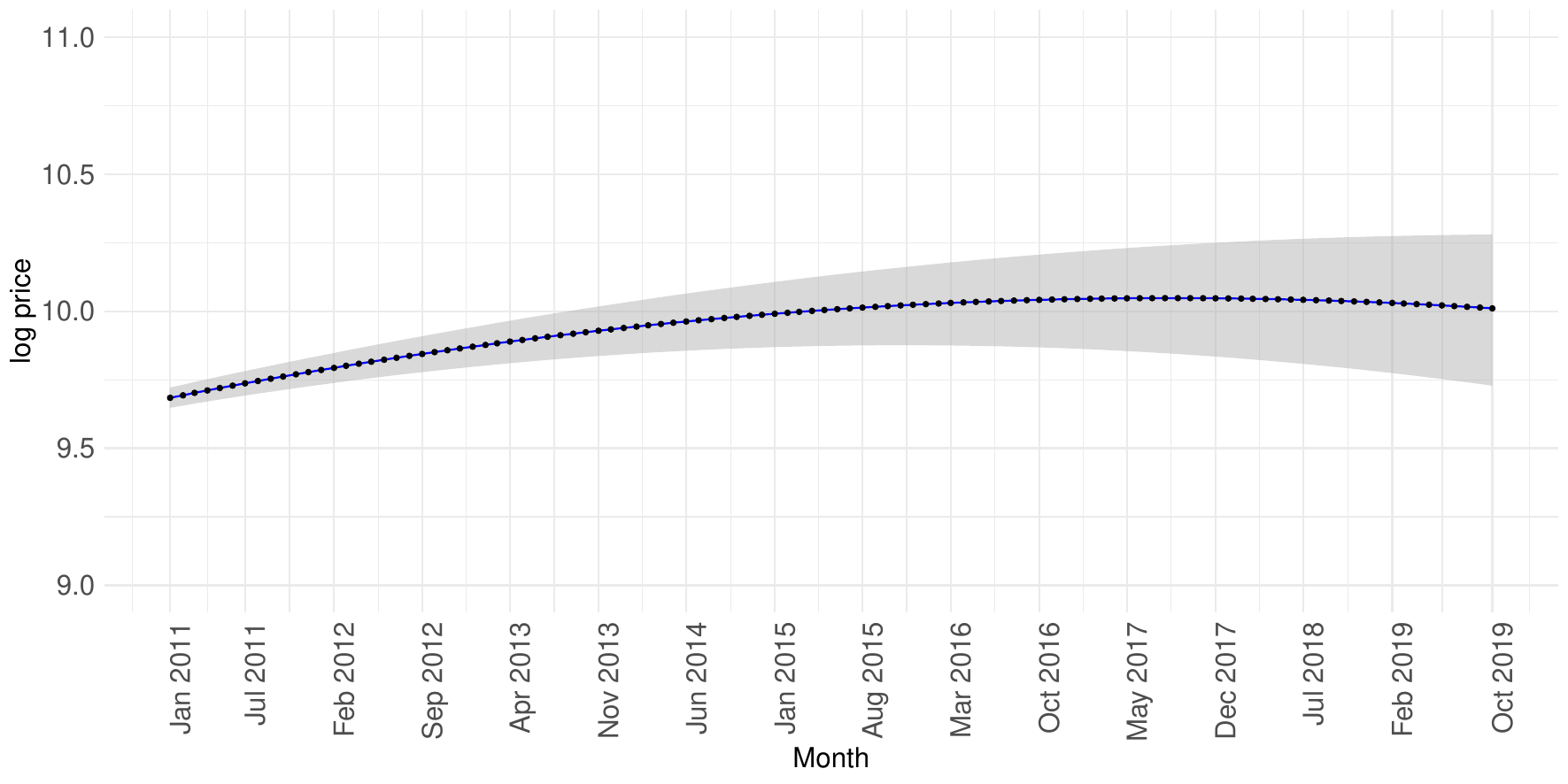} %24_may
    \caption{Average trend of the house price (log-transformed) for all locations over 106 time.}% The mean structure was utilised to model the log price over time-points.}
    \label{fig:interval}
\end{figure}

First, we notice that both the linear and quadratic terms of the trend coefficients are significant, with the linear coefficient being positive and the quadratic coefficient being negative. Moreover, the magnitude of the linear trend coefficient is quite higher than the other one, indicating that the house price is increasing over time. To demonstrate the impact of time trend, we utilize the intercept and time trend coefficients to generate fitted mean log price values for all MSOAs across the 106 time-points. As shown in \Cref{fig:interval}, it depicts a curvilinear relationship between the time trend and the response variable. The average log price per square meter increases at a diminishing rate over time and eventually flattens out. This trend is akin to the concept of diminishing marginal returns in economics, where the later time periods produce smaller increases in the variable of interest. These results are consistent with a study by \cite{ons}, where the recent trends in house price growth in London was analyzed using the UK House Price Index from January 2012 to July 2018. The study found that the growth in house prices had declined since 2016.

Turning attention to the estimates of the covariates, detailed interpretations of their impact on the response variable are discussed next.

\subsubsection{Effect of area and number of rooms}

Our analysis unveils a straightforward relationship between property size and price in London. The negative coefficient associated with log area indicates that as a property's size increases, its value per square meter tends to decrease. This observation is in line with common expectations and matches the results reported in extant literature, e.g., \cite{chegut2016energy}. 

Furthermore, the number of rooms within a property emerges as a significant factor. Properties with more than two rooms command notably higher prices compared to those with 1 or 2 rooms. This effect becomes more pronounced for properties with over five rooms, where the impact on price per square meter is most substantial. It highlights the premium placed on larger, more spacious homes in the London region, reflecting the diverse housing preferences of potential buyers.

In summary, larger properties with same number of rooms tend to have lower prices per square meter than smaller properties. This finding aligns with the results reported in \cite{fuerst2015does}.

\subsubsection{Effect of property characteristics}

The type of property stands out as a pivotal factor influencing prices. We find that terraced and flat properties tend to have lower prices per square meter compared to semi-detached houses. Strikingly, flats are at the lower end of the price spectrum, while detached houses command the highest prices. This finding aligns with the study conducted by \cite{fuerst2015does}, where terraced properties were considered the base category, and both detached and semi-detached properties were found to be on the premium side, while flats were on the lower end. This result underscores the importance of property type as a driver of price volatility in the London real estate market and reflects the diverse architectural landscape of the region. 

The presence of fireplaces within properties is strongly associated with higher prices. It aligns with the perception of fireplaces as desirable amenities for homeowners, contributing positively to property values, and is consistent with the observations of \cite{zhang2016flood}. Meanwhile, we find that the ventilation systems exert a discernible influence on house prices. Properties equipped with extract mechanical ventilation and both supply and extract mechanical ventilation systems tend to exhibit higher prices compared to those reliant on natural ventilation. One may connect this to the growing emphasis on indoor air quality and environmental factors among property buyers. It reinforces the desirability of properties that offer superior air quality, facilitated by mechanical ventilation systems.

% Houses equipped with fireplaces tend to command higher prices than houses without a fireplace. This finding underscores the enduring appeal of traditional, cozy features that enhance the overall desirability and market value of a property. , who found that fireplace equipment has greater value in higher-priced homes

%Particularly noteworthy is the finding that properties featuring both supply and extract ventilation systems command the highest prices within this category, highlighting the premium placed on comprehensive ventilation solutions in the real estate market.

\subsubsection{Effect of carbon emission} % and their evolving impact on house prices}

Another intriguing aspect of our analysis is the investigation of the relationship between carbon emissions and house prices. The increasing need of a cleaner environment is likely to impact the economy of real estate and we attempt to capture the evolving dynamics of sustainability in the London housing market through this analysis. By considering appropriate regressors, we delve into the multifaceted aspects of this relationship, including its impact, interaction with time, and the implications for supply and demand.

\textbf{Impact of carbon emission on house price}: Our findings highlight the significant impact of carbon emissions on house prices. We observe a negative relationship, indicating that properties with lower carbon emission values tend to command higher prices. This emphasizes the growing importance of sustainability and environmental consciousness among real estate buyers. It suggests that buyers are increasingly willing to invest more in homes with a reduced carbon footprint, reflecting the evolving trend toward eco-friendly living spaces. The findings align with those of \cite{gerassimenko2023impact}, who noted that energy-efficient properties, characterized by low carbon emissions, tend to command a price premium.

\textbf{The interaction with time}: A more nuanced observation emerges when we consider the interaction between carbon emissions and time. Over time, we observe a decline in willingness to pay premium for lower carbon emissions, reflecting a shift in buyer behavior. This observation is supported by the positive coefficient associated with the interaction term. In practice, while carbon emissions continue to be an important factor for buyers, their impact on house prices has gradually diminished over the years. This trend raises interesting questions about the evolution of buyer preferences and the broader real estate market. %It implies that as sustainability practices become more mainstream, the market is adjusting to accommodate eco-friendly options at more equitable price points. It could be attributed to increased supply of sustainable properties as builders respond to demand or to growing consumer awareness and acceptance of eco-friendly living. In either case, it signifies a market that is in flux, adapting to the changing landscape of sustainability.

\textbf{Supply and demand implications}: Finally, the interplay between carbon emissions and house prices also has implications for supply and demand dynamics. As sustainability becomes a central consideration for buyers, properties with lower carbon emissions are likely to see increased demand. This heightened demand may prompt the builders to focus on constructing more environmentally friendly properties, further influencing the supply side of the equation. However, our observation that the price differential between eco-friendly and conventional properties is diminishing suggests that the market is finding a balance. As supply increases and the premium on sustainability lessens, a broader range of buyers can access these eco-friendly properties \citep[see][for pertinent discussions]{ma2017impacts, blanco2023knocking}. This equilibrium benefits both buyers and the environment.

\subsection{Comparison between the fits of different models}

As mentioned in \Cref{sec:model_matrices}, we conduct a comprehensive comparative study to assess the fitting and accuracy of different models. \Cref{tab:compare} displays the accuracy metrics on overall data to evaluate the goodness of fit. The results unequivocally demonstrate that our proposed model outperforms the other models, indicating its superior ability to capture the spatial and temporal dependencies in the data and achieve a better fit. Notably, D\&C-STBD explains the highest variation, while the temporal model falls short in explaining the variance effectively. We also observe that the D\&C-SBD method requires significantly less time compared to the spatio-temporal model or the spatial hedonic model, yet it provides a great fit. This highlights the efficiency and accuracy of the spatial component in the method, especially in the context of the divide-and-conquer approach.

\begin{table}[hbt!]
\centering
\caption{Comparison between different models for fitting the entire data.}
  \renewcommand{\arraystretch}{1.2}
 \label{tab:compare}
\begin{tabular}{ccccccc}
  \hline
 Model& $R^2$ & MAE & MAPE & RMSE & VAE & Computation time (in mins) \\ 
  \hline
 Simple hedonic  &0.166 & 0.327 &3.821\% & 0.426& 0.075&2.19\\
 Spatial hedonic &0.605&0.218 &2.568\% &0.293 & 0.038&4.92\\
 Temporal & 0.173 &0.325 &3.80\% &0.424 & 0.074&25.23\\
 SBD &0.744 &0.169 &1.984\% &0.236 &0.027 &60 \\
 D\&C-SBD & 0.739 &  0.171 & 2.013\% & 0.238 & 0.028 & 1.50\\
  D\&C-STBD  & 0.796 & 0.150 & 1.767\% &0.211 & 0.022 & 540\\
   \hline
\end{tabular}
\end{table}

It is worth highlighting that, in an attempt to understand the efficacy of the divide-and-conquer step, we tried to directly implement the model used in D\&C-SBD. When comparing the performance of the two approaches, we find that the method without the D\&C step achieved only a marginally better fit ($R^2$ of 0.744, as compared to 0.739 in D\&C-SBD). However, this improvement came at a significant computational cost, with the spatial model taking approximately 40 times longer. Due to memory constraints, we could not perform the complete spatio-temporal model on the entire dataset. Nevertheless, our findings indicate that the proposed model provides approximate results closely aligned with what the actual model would produce. We acknowledge that our proposed model requires more time due to its incorporation of both spatial and temporal dimensions. However, this additional computational effort is justified by the model's ability to capture relationships more effectively compared to other models. 

To further support the above, it is important to recognize that the hedonic or spatial hedonic model offer quite different results in terms of the coefficient estimates for some parameters. For instance, contrary to the common knowledge and existing literature, the hedonic model identifies a significantly positive impact on the price if the property type is flat and these models fail to account for the significant premium associated with more spacious homes. These inconsistencies underscores the intricate nature of housing dynamics and emphasizes the necessity for sophisticated analysis beyond conventional linear models. As we explained in the previous subsection, suitable specification of spatial and temporal dependence in our model helps in capturing the effect sizes in an appropriate way, and renders valuable insights about the real estate markets. Furthermore, the proposed methodology offers a superior balance between computational efficiency and accuracy, and can be a valuable option for large datasets with spatial and temporal relationships.

\subsection{Error parameters and residual diagnostics}

We now focus on the estimated error structures in the proposed model and aim to diagnose the residuals to assess whether the model has appropriately captured the dependence patterns in the dataset. The estimated parameters of the Gaussian processes governing the residuals show that the spatio-temporal error process explains more variation in the data than the white noise, as the estimated $\sigma_v^2$ is almost double in magnitude compared to $\hat\sigma_\epsilon^2$. This results in a significant contribution of the $v(s,t)$ process to the overall variance and endorses the use of a spatio-temporal model to enhance our understanding of London's house price dynamics. Furthermore, we note that the estimated decay parameters $\hat\phi_s = 2.402$ and $\hat\phi_t = 0.528$ reveal practical implications. With the equation $\exp(-\phi d) \approx 0.05$ in mind, we can argue that the estimated values of these decay parameters imply that the spatial correlation remains significant for up to a distance of 1.25 kilometers, while the temporal dependence remain significant over a period of 5-6 months.

\begin{figure}[!htb]
    \centering
    \includegraphics[width = 0.7\textwidth,keepaspectratio]{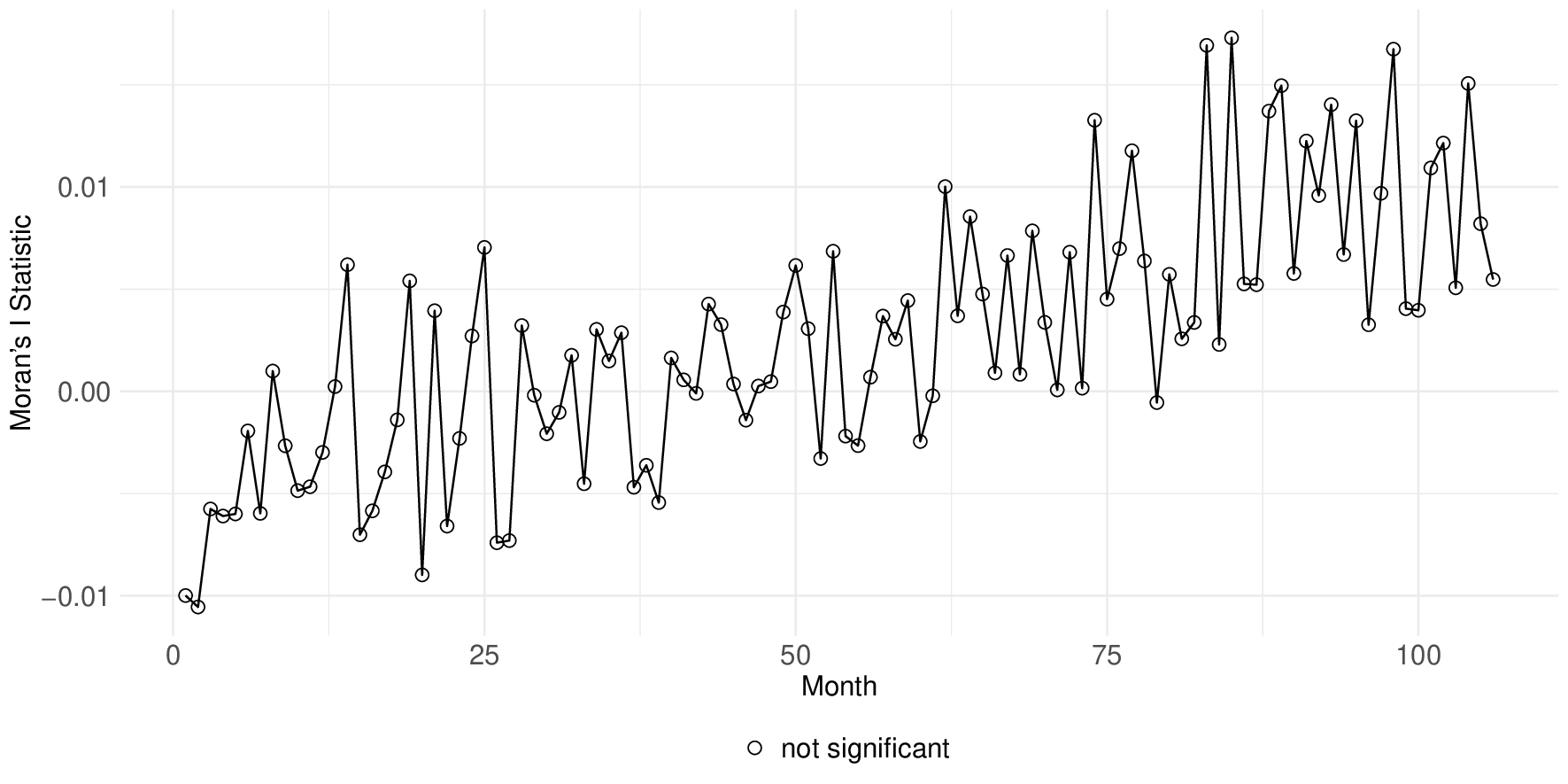}
    %moran_result_plot
    \caption{Moran’s I statistic for the residuals for all the monthly time intervals in the entire London region. Black circle indicates a significant 
spatial correlation at 1\% level of significance.}
    \label{fig:moran_residual}
\end{figure}

We next look at a few diagnostic tests to assess the proposed model's adequacy in capturing spatial and temporal dependence patterns. In the same spirit as in \Cref{sec:data}, we compute the Moran's I statistics for the residuals at all time-points following \eqref{eqn : moran}. The plot is presented in \Cref{fig:moran_residual}, and it evidently indicates that spatial correlations are no longer significant at 1\% level. Contrasting it with \Cref{fig:moran}, we can ascertain that the spatial correlation has been successfully captured by our method. 

\begin{figure}[!htb]
    \centering
    \includegraphics[width = 0.7\textwidth]{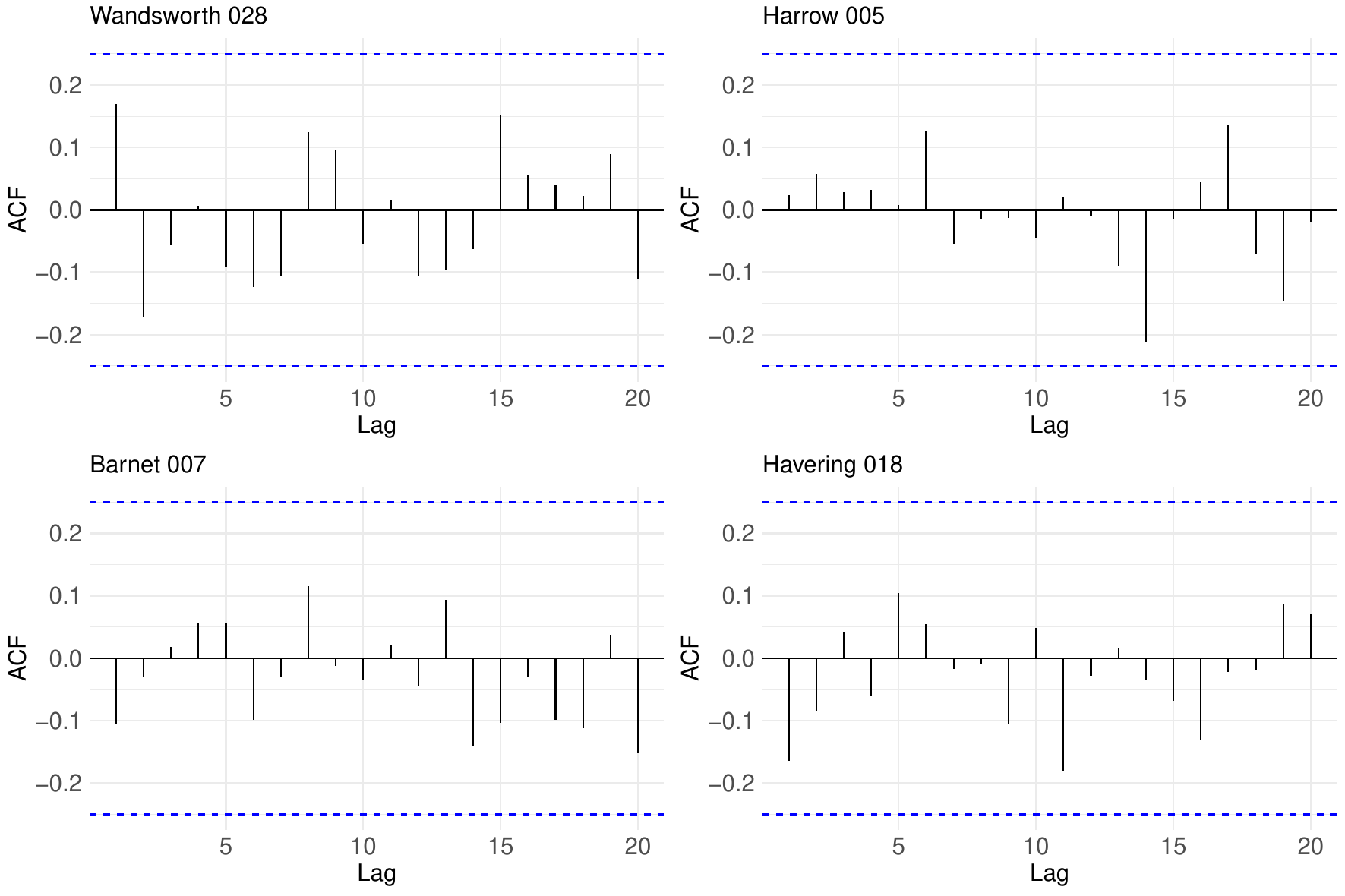}
    \caption{ACF plots for the residuals for four randomly selected MSOAs. The dotted lines are corresponding to the critical value at 1\% significance level.}
    %acf_uk
    \label{fig:ACF_residual}
\end{figure}

For identifying any remaining temporal correlation, we analyze ACF for the residuals across different locations. We find that the residual series for majority of the MSOAs do not have significant temporal autocorrelation. For instance, the ACF plots of residuals for the same four MSOAs as before are illustrated in \Cref{fig:ACF_residual}, which signifies that the temporal correlation does not exist anymore.

In summary, our proposed model effectively captures both spatial and temporal dependence patterns, providing a good fit for the data. In addition, we also looked at the histogram and the quantile-quantile (QQ) plot of the residuals, to assess the normality assumption. The analysis indicates that the residuals have a symmetric distribution around 0, with a hint of a heavier tail than a normal distribution. It might be suggestive of a non-Gaussian residual distribution, and exploring this further could be a valuable direction for future research. We discuss this in \Cref{sec:conclusion}.

\subsection{Predictive performance}
\label{subsec:prediction}

% After discussing the effectiveness of all the models to capture the relationship in better way and how well our proposed model captured both spatial and temporal dependency, we leverage our proposed model to predict future house prices, both for new locations and existing ones. This predictive capacity represents a key advantage of our approach, which we explore further in the subsequent subsection.

As a final exploration, we assess our model's predictive capabilities in two scenarios, looking at a future horizon of one year in both cases. In the first scenario, the training set comprises 94 monthly observations for 983 MSOAs, totaling 597,172 sample observations, while the test set includes the last 12 monthly observations (November 2018 to October 2019) for these MSOAs, totaling 54,030 samples. In the second scenario, our objective is to assess the predictive accuracy for unobserved MSOAs. To that end, we consider 23 randomly selected MSOAs. The training set includes the initial 94 time points for the remaining 960 MSOAs, totaling 584,672 samples. The test set in this case comprises 1,068 samples from the 23 selected MSOAs over the last 12 time points.

For both scenarios, we employ a similar division strategy to the one used when fitting the entire dataset. In the first case, the 983 MSOAs are divided into 20  subsets, with 19 subsets containing 49 MSOAs each, and the $20^{th}$ subset keeping 52 locations. In the second scenario, we split the 960 MSOAs into 20 subsets, with each subset containing 48 MSOAs. We fit the proposed model using the method described in \Cref{subsec : combine}, obtain the required parameter estimates, and make predictions for each location and time in the test set using the procedure outlined in \Cref{subsec : pred}. 

\begin{figure}[!htb]
\vspace{-2cm}
    \centering
    \includegraphics[width = 0.7\textwidth,keepaspectratio]{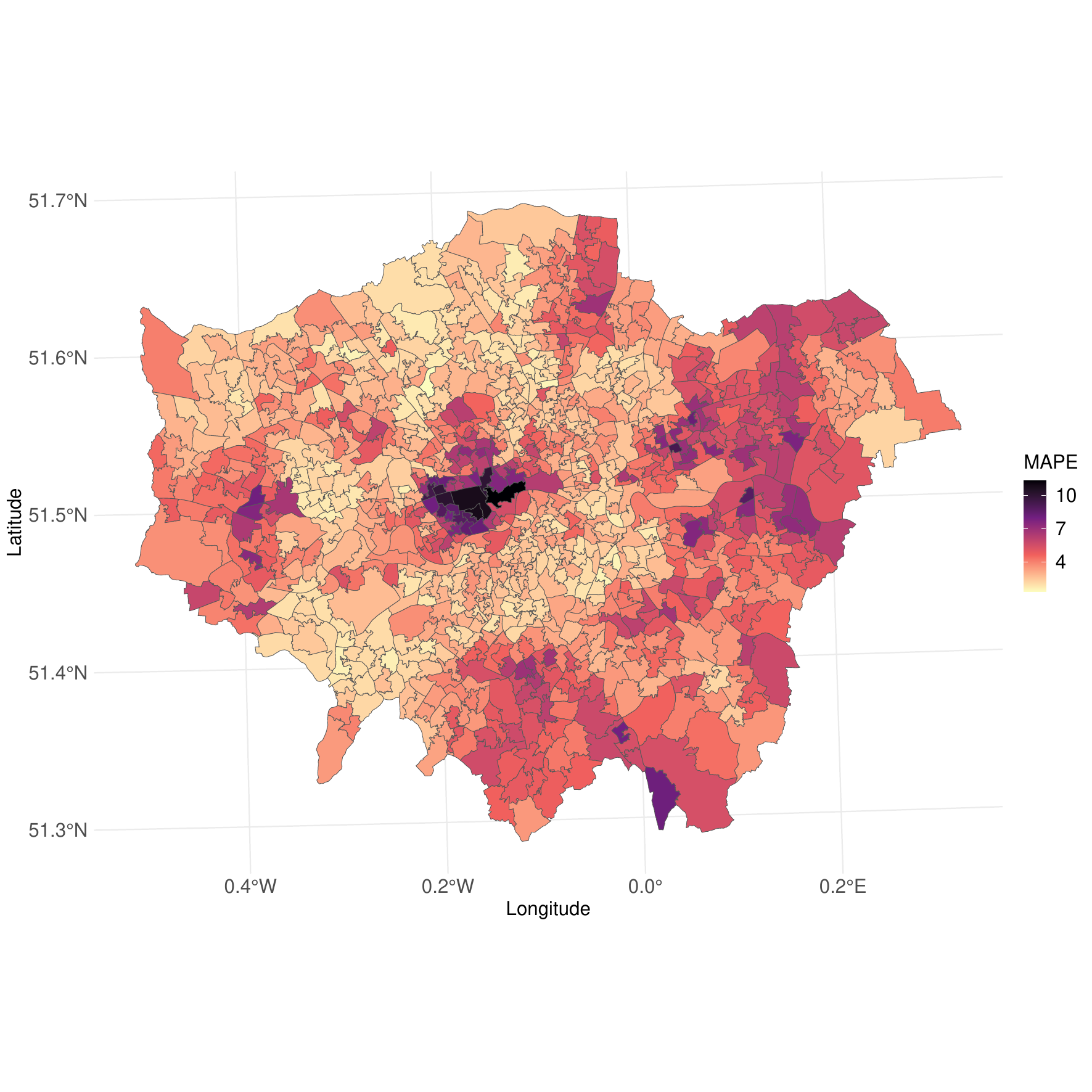}
  \vspace{-2cm}\captionsetup{justification=centering} 
    \caption{Prediction MAPE (in \%) in log price per square meter for all  MSOAs for 12 months in the first case.}
    \label{fig:london_price_case1}
\end{figure}

First, to understand the predictive ability in the spatial aspect, we calculate the Mean Absolute Percentage Error (MAPE) for all MSOAs in the first scenario, where the average is taken over the 12 time-points in the test set. These values are displayed in \Cref{fig:london_price_case1}. There, it can be observed that the MAPE varies across MSOAs, ranging from 1.3\% to 11.25\%. Notably, only four MSOAs, namely ``Westminster 018'', ``Kensington and Chelsea 012'', ``Westminster 019'' and , ``Kensington and Chelsea 008'' exhibit MAPE values exceeding 10\%. This indicates potential areas where our predictive models may require further refinement.

To gain a broader perspective, we delve into local authority (region) wise MAPE, aiming to discern trends in error rates across different regions. Our investigation reveals that region-wise MAPE spans from 2.3\% to 8.3\%. Particularly noteworthy is the highest MAPE observed in the ``Kensington and Chelsea'' region, contrasted  with the lowest MAPE in the ``Kingston upon Thames'' region. Conversely, for the remaining regions, MAPE values remain below 7\%, suggesting relatively lower prediction errors. These discrepancy could potentially be attributed to market instability in the UK real estate sector during the corresponding period in 2018 and 2019. In an attempt to gain deeper insights, a detailed examination of property-wise errors is conducted next. In the first scenario, the test data contains a total of 54,030 house sales. We calculate the Absolute Percentage Error (APE) in forecasting the price for each transaction and categorize them into different APE brackets, as detailed under Case 1 in \Cref{tab:APE}. It shows that about 75\% of properties have APE of less than 5\%, while only about 3.5\% of properties exceed an APE of 10\%. These results underscore the effectiveness of our proposed approach in providing accurate predictions at the individual property level. The findings are identical in the second case as well. % The analysis reveals that approximately 77\% of properties exhibit an APE of less than 5\%, while merely 1.7\% surpass an APE of 10\%. These findings underscore the efficacy of our proposed approach in furnishing precise predictions for unobserved locations at the granularity of individual properties.  % and indicate strong predictive performance in terms of property location. % This allows us to assess the distribution of prediction errors throughout the dataset, providing valuable insights into our model's predictive performance.

\begin{table}[hbt!]
\centering
\caption{Distribution of houses corresponding to different levels of absolute percentage errors for both cases.}
\renewcommand{\arraystretch}{1.2}
\label{tab:APE}
\begin{tabular}{lcccc}
\hline
\multicolumn{1}{c}{} & \multicolumn{2}{c}{Case 1 (total sample size: 54030)} & \multicolumn{2}{c}{Case 2 (total sample size: 1068)} \\
\cline{1-3} \cline{4-5}
APE & Number of cases & Percentage of cases  & Number of cases & Percentage of cases \\
\hline
Less than 3\% & 25604 & 47.39\%   & 508 & 47.57\% \\
3 to 5\% & 14267 & 26.41\% & 306 & 28.65\% \\
5 to 10\% & 12238 & 22.65\% &  236 & 22.10\% \\
More than 10\% & 1921 & 3.55\% &  18 & 1.69\% \\
\hline
\end{tabular}
\end{table}

\begin{figure}[!htb]
    \centering
    \includegraphics[width = 0.8\textwidth,keepaspectratio]{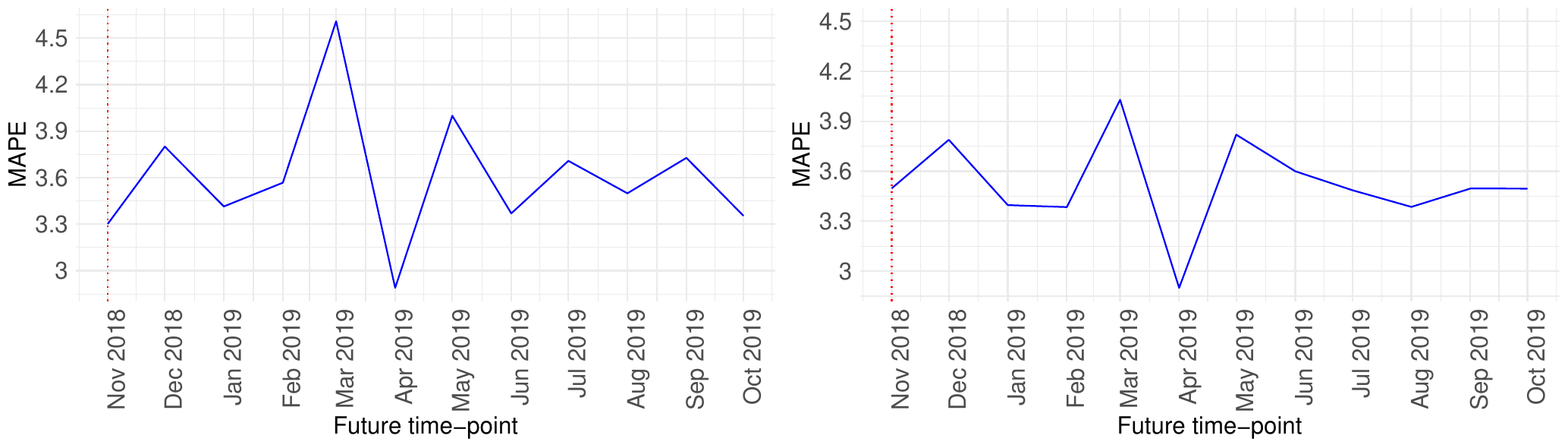}
    \caption{Prediction MAPE (in \%) for 23 locations: left panel corresponds to the case where the locations are included in the training set, right panel corresponds to the case where they are not part of the training set.}
    \label{fig:london_price_time}
\end{figure}

Next, focus on the second scenario with the 23 locations unobserved in the training data. In \Cref{fig:london_price_time}, we present the MAPE for these locations according to different time-points in the future. It helps us quantify the predictive performance over time for unobserved locations. In the figure, the right panel corresponds to this whereas in the left panel, we include the accuracy for the scenario where these locations are present in the training data. Interestingly, in both scenarios, the MAPE ranges between 2.9\% and 4.6\%, which confirms that the proposed method can offer good predictive certainty for properties from new locations as well. Moreover, the results showcase a consistently accurate level of forecasting for the response variable over all time-points in future. Finally, we present the individual MAPE values for the 23 MSOAs across the 12 time points in \Cref{tab:MAPE_tab}. Our analysis reveals that, barring three specific instances involving MSOAs and time points, we have accurately replicated the original log house prices, achieving a MAPE of less than 10\%.

% \update{In addition, we compute the Absolute Percentage Error (APE) similar to case 1 for forecasting transaction prices, as elucidated under case 2 in  \Cref{tab:APE}. Additionally, we provide an in-depth discussion of Mean Absolute Percentage Error (MAPE) by presenting the MAPE values for all 23 MSOAs across the forthcoming 12 time points in \Cref{tab:MAPE_tab}. Our analysis reveals that, barring three specific instances involving MSOAs and time points, we have accurately replicated the original log house prices, achieving a MAPE of less than 10\%.}

\begin{table}[hbt!]
\centering
\caption{MAPE values calculated for the 23 out-of-sample MSOAs across all 12 future time points, which were not part of the training dataset, employing the proposed model. Missing cells indicate that there was no sale in those months. More than 10\% errors are marked in bold.}
\label{tab:MAPE_tab}
{\small
\begin{tabular}{lcccccccccccc}
  \hline
  MSOA & Nov & Dec & Jan & Feb & Mar & Apr & May & Jun & Jul & Aug & Sep & Oct \\ 
 & 2018 & 2018 & 2019 & 2019 & 2019 &2019 &2019 &2019 &2019 &2019 &2019 &2019 \\
  \hline
  Barnet 027 & 2.88 & 1.01 & 1.93 & 3.25 & 1.38 & 2.11 & 7.14 & 1.69 & 1.13 & 2.02 & 5.30 & 2.79 \\ 
  Barnet 035 & 1.11 & 2.45 & 1.02 & 1.98 & 2.76 & 3.14 & 3.81 & 4.62 & 2.74 & 2.19 & 0.60 & 1.88 \\ 
  Brent 007 & 1.91 & 3.03 & 0.67 & 2.62 & 2.21 & 1.88 & 4.25 & 4.54 & 2.72 & 2.49 & 3.59 & 5.01 \\ 
  Brent 009 & 1.65 & 2.82 & 1.96 & 1.84 & 9.08 & 2.00 &  & 2.06 & 2.92 & 2.18 & 6.71 &  \\ 
  Brent 013 &  & 4.60 & 6.51 & 4.84 & 8.21 & 1.43 & 3.30 & 1.68 & 4.23 & 5.50 & 3.16 & 3.28 \\ 
  Brent 026 & 2.84 & 0.36 & 8.11 & 0.38 & 1.54 &  & 5.28 & 2.14 & 2.45 & 2.01 & 0.81 & 5.15 \\ 
  Croydon 021 & 7.33 & 5.87 & 5.21 & 4.54 & 6.23 & 3.51 & 5.91 & 3.70 & 4.73 & 4.89 & 4.53 & 6.58 \\ 
  Ealing 030 & 1.49 & 4.02 & 0.72 & 2.60 & 0.11 & 2.03 & 0.83 & 1.55 & 1.09 & 1.64 & 2.31 & 2.05 \\ 
  Hackney 024 & 3.64 & 2.22 & 4.70 & 4.92 & 3.69 & 1.99 & 1.21 & 2.59 & 4.67 & 3.85 & 1.63 & 3.53 \\ 
   Hammersmith 024 & 4.83 & 5.02 & 3.01 & 4.23 & 3.99 & 4.66 & 3.37 & 7.08 & 5.99 & 4.79 & 3.37 & 4.01 \\ 
   Harrow 019 & 5.56 & 3.61 & 1.02 & 1.40 & 2.00 & 1.34 & 2.59 & \textbf{20.47} & 2.35 & 2.08 & 2.03 &  \\ 
   Harrow 024 & 2.37 & 2.50 & 2.26 & 2.92 & 4.21 & 4.30 & 3.87 & 1.83 & 4.13 & 1.77 & 2.79 & 5.95 \\ 
   Islington 008 & 4.60 & 6.73 & 4.55 & 2.89 & 2.12 & 1.86 & 5.75 & 3.94 & 3.04 & 3.76 & 0.03 & 4.74 \\ 
   Islington 019 & 2.58 &  & 3.23 & 4.09 & 1.57 & 2.12 & 1.78 & 2.45 & 3.47 & 3.01 & 1.81 & 4.77 \\ 
   Lambeth 022 & 3.10 & 0.85 & 2.35 & 3.92 & 2.65 & 1.60 & 0.97 & 0.93 & 2.27 & 2.53 & 2.97 & 1.43 \\ 
   Merton 010 & 2.69 & 2.73 & 1.83 & 2.10 & 3.57 & 1.68 & 0.76 & 2.88 & 2.24 & 2.96 & 2.28 & 1.93 \\ 
   Merton 011 & 1.54 & 1.91 & 2.31 & 2.20 & 2.00 & 1.51 & 2.11 & 1.62 & 1.62 & 2.25 & 1.61 & 1.87 \\ 
   Merton 013 & 3.45 & 3.83 & 3.43 & 5.90 & 4.68 & 3.86 & 4.33 & 3.33 & 1.47 & 2.72 & 4.94 & 3.88 \\ 
   Newham 033 & 2.15 &  & 2.96 & 4.57 &  &  & 2.60 & 3.03 & 2.43 & 3.20 & 0.82 &  \\ 
   Southwark 002 & 4.73 & 4.70 & 2.44 & 2.89 & 4.57 & 2.84 & 4.61 & 3.40 & 4.12 & 3.60 & 2.57 & \textbf{10.93} \\ 
   Sutton 011 & 3.30 & 4.08 & 3.92 & 3.92 & 4.54 & 3.87 & 5.52 & 5.88 & 4.38 & 4.89 & 4.12 & 3.91 \\ 
   Sutton 019 & 4.51 & 5.26 & 1.42 & 4.41 & 5.21 & 4.68 & 4.52 & 5.03 & 5.28 & 4.96 & \textbf{11.87} &  \\ 
   Westminster 016 & 7.10 & 7.21 & 8.21 & 5.02 & 7.33 & 4.90 & 7.00 & 6.45 & 5.82 & 5.23 & 6.27 &  \\ 
   \hline
\end{tabular}}
\end{table}

\section{Conclusion}
\label{sec:conclusion}

In this study, we have introduced a novel divide-and-conquer spatio-temporal modeling approach specially designed for large datasets. One distinctive aspect of our proposed methodology is its ability to accommodate multiple observations for the same location at the same time-points, as well as handling instances of zero transactions or missing data within the modeling framework. The model is applied to the house price data from London. The dataset comprises 106 monthly observations from 983 MSOAs. Through exploratory analysis, we identify significant spatial and temporal correlations, leading us to propose a spatio-temporal model. Built upon a Bayesian framework, our modeling process utilized Gibbs and 
discrete slice samplers for parameter estimation, allowing us to gain several key insights into the factors influencing house prices in London. Carbon emissions significantly influence property prices, reflecting the growing importance of sustainability. However, this influence is gradually diminishing, prompting a market shift. Supply and demand dynamics are adapting, with eco-friendly construction becoming more viable. Property characteristics, such as type and amenities, also affect prices. Our proposed model elucidates market complexity, emphasizing the need for adaptability to changing buyer preferences and sustainability trends in London's real estate sector.

Before concluding the paper, we want to suggest a few directions for future research. Exploratory data analysis revealed variations in the time trend effect on house prices among different MSOAs. An exciting avenue for future investigation involves exploring a more generalized modeling framework capable of accounting for regionally varying coefficients for covariates alongside the spatio-temporal error process. This extension could help identify specific regions experiencing significant slowdowns in growth rates or areas where people are willing to pay more for lower carbon emissions. In the same spirit, imposing a region-wise heteroskedastic structure on the white noise can improve the fit as well as predictive capability of the model. One may also relax the Gaussianity assumption and extend the model to heavy-tailed distributions.

From an application standpoint, our method's computational efficiency enhances its utility, making it an advantageous tool for a wide range of practical situations. The proposed approach extends beyond the confines of London's housing market, becoming a versatile tool applicable to various other problems, thus providing value for stakeholders and policymakers across different domains. For instance, it can be effectively adapted to analyze space-time characteristics in datasets where high-level granularity is available with minimal missingness. A few examples can be found in environmental research (e.g., air pollution, rainfall analysis), epidemiology (e.g., COVID-19 data) or other economic applications (e.g., agricultural production).

%\section*{Data availability statement}

%This dataset is publicly available in ``UK Data Service reshare'' at \url{https://reshare.ukdataservice.ac.uk/854240/}. 

%\section*{Acknowledgements}

%under the Open Government Licence v3.0.

\section*{Declaration of interest}

The authors declare no conflict of interest.

\bibliography{references}

\begin{thebibliography}{57}
\expandafter\ifx\csname natexlab\endcsname\relax\def\natexlab#1{#1}\fi
\providecommand{\url}[1]{\texttt{#1}}
\providecommand{\href}[2]{#2}
\providecommand{\path}[1]{#1}
\providecommand{\DOIprefix}{doi:}
\providecommand{\ArXivprefix}{arXiv:}
\providecommand{\URLprefix}{URL: }
\providecommand{\Pubmedprefix}{pmid:}
\providecommand{\doi}[1]{\href{http://dx.doi.org/#1}{\path{#1}}}
\providecommand{\Pubmed}[1]{\href{pmid:#1}{\path{#1}}}
\providecommand{\bibinfo}[2]{#2}
\ifx\xfnm\relax \def\xfnm[#1]{\unskip,\space#1}\fi
%Type = Article
\bibitem[{Arribas-Bel(2014)}]{arribas2014accidental}
\bibinfo{author}{Arribas-Bel, D.}, \bibinfo{year}{2014}.
\newblock \bibinfo{title}{Accidental, open and everywhere: Emerging data
  sources for the understanding of cities}.
\newblock \bibinfo{journal}{Applied Geography} \bibinfo{volume}{49},
  \bibinfo{pages}{45--53}.
%Type = Book
\bibitem[{Banerjee et~al.(2014)Banerjee, Carlin and
  Gelfand}]{banerjee2014hierarchical}
\bibinfo{author}{Banerjee, S.}, \bibinfo{author}{Carlin, B.P.},
  \bibinfo{author}{Gelfand, A.E.}, \bibinfo{year}{2014}.
\newblock \bibinfo{title}{Hierarchical Modeling and Analysis for Spatial Data}.
\newblock \bibinfo{publisher}{CRC Press}.
%Type = Article
\bibitem[{Beamonte et~al.(2010)Beamonte, Gargallo and
  Salvador}]{beamonte2010analysis}
\bibinfo{author}{Beamonte, A.}, \bibinfo{author}{Gargallo, P.},
  \bibinfo{author}{Salvador, M.}, \bibinfo{year}{2010}.
\newblock \bibinfo{title}{{Analysis of housing price by means of STAR models
  with neighbourhood effects: a Bayesian approach}}.
\newblock \bibinfo{journal}{Journal of Geographical Systems}
  \bibinfo{volume}{12}, \bibinfo{pages}{227--240}.
%Type = Article
\bibitem[{Bickel and Freedman(1981)}]{bickel1981some}
\bibinfo{author}{Bickel, P.J.}, \bibinfo{author}{Freedman, D.A.},
  \bibinfo{year}{1981}.
\newblock \bibinfo{title}{Some asymptotic theory for the bootstrap}.
\newblock \bibinfo{journal}{The annals of statistics} \bibinfo{volume}{9},
  \bibinfo{pages}{1196--1217}.
%Type = Techreport
\bibitem[{Blanco and Neri(2023)}]{blanco2023knocking}
\bibinfo{author}{Blanco, H.}, \bibinfo{author}{Neri, L.}, \bibinfo{year}{2023}.
\newblock \bibinfo{title}{Knocking It Down and Mixing It Up: The Impact of
  Public Housing Regenerations}.
\newblock \bibinfo{type}{Technical Report}. IZA Discussion Papers.
%Type = Article
\bibitem[{Blatt et~al.(2023)Blatt, Chaudhuri and Manner}]{blatt2023changepoint}
\bibinfo{author}{Blatt, D.}, \bibinfo{author}{Chaudhuri, K.},
  \bibinfo{author}{Manner, H.}, \bibinfo{year}{2023}.
\newblock \bibinfo{title}{A changepoint analysis of uk house price spillovers}.
\newblock \bibinfo{journal}{Regional Studies} \bibinfo{volume}{57},
  \bibinfo{pages}{1223--1238}.
%Type = Article
\bibitem[{Botchkarev(2019)}]{botchkarev2019new}
\bibinfo{author}{Botchkarev, A.}, \bibinfo{year}{2019}.
\newblock \bibinfo{title}{A new typology design of performance metrics to
  measure errors in machine learning regression algorithms}.
\newblock \bibinfo{journal}{Interdisciplinary Journal of Information,
  Knowledge, and Management} \bibinfo{volume}{14}, \bibinfo{pages}{045--076}.
%Type = Article
\bibitem[{Can(1990)}]{can1990measurement}
\bibinfo{author}{Can, A.}, \bibinfo{year}{1990}.
\newblock \bibinfo{title}{The measurement of neighborhood dynamics in urban
  house prices}.
\newblock \bibinfo{journal}{Economic geography} \bibinfo{volume}{66},
  \bibinfo{pages}{254--272}.
%Type = Article
\bibitem[{Chegut et~al.(2016)Chegut, Eichholtz and
  Holtermans}]{chegut2016energy}
\bibinfo{author}{Chegut, A.}, \bibinfo{author}{Eichholtz, P.},
  \bibinfo{author}{Holtermans, R.}, \bibinfo{year}{2016}.
\newblock \bibinfo{title}{Energy efficiency and economic value in affordable
  housing}.
\newblock \bibinfo{journal}{Energy Policy} \bibinfo{volume}{97},
  \bibinfo{pages}{39--49}.
%Type = Article
\bibitem[{Chi et~al.(2021a)Chi, Dennett, Ol{\'e}ron-Evans and
  Morphet}]{chi2021new}
\bibinfo{author}{Chi, B.}, \bibinfo{author}{Dennett, A.},
  \bibinfo{author}{Ol{\'e}ron-Evans, T.}, \bibinfo{author}{Morphet, R.},
  \bibinfo{year}{2021}a.
\newblock \bibinfo{title}{A new attribute-linked residential property price
  dataset for england and wales, 2011 to 2019}.
\newblock \bibinfo{journal}{UCL Open: Environment Preprint} .
%Type = Article
\bibitem[{Chi et~al.(2021b)Chi, Dennett, Ol{\'e}ron-Evans and
  Morphet}]{chi2021shedding}
\bibinfo{author}{Chi, B.}, \bibinfo{author}{Dennett, A.},
  \bibinfo{author}{Ol{\'e}ron-Evans, T.}, \bibinfo{author}{Morphet, R.},
  \bibinfo{year}{2021}b.
\newblock \bibinfo{title}{Shedding new light on residential property price
  variation in england: A multi-scale exploration}.
\newblock \bibinfo{journal}{Environment and Planning B: Urban Analytics and
  City Science} \bibinfo{volume}{48}, \bibinfo{pages}{1895--1911}.
%Type = Article
\bibitem[{Chi et~al.(2022)Chi, Dennett, Ol{\'e}ron-Evans and
  Morphet}]{chi2022delineating}
\bibinfo{author}{Chi, B.}, \bibinfo{author}{Dennett, A.},
  \bibinfo{author}{Ol{\'e}ron-Evans, T.}, \bibinfo{author}{Morphet, R.},
  \bibinfo{year}{2022}.
\newblock \bibinfo{title}{Delineating the spatio-temporal pattern of house
  price variation by local authority in england: 2009 to 2016}.
\newblock \bibinfo{journal}{Geographical Analysis} \bibinfo{volume}{54},
  \bibinfo{pages}{219--238}.
%Type = Article
\bibitem[{Cook and Watson(2016)}]{cook2016new}
\bibinfo{author}{Cook, S.}, \bibinfo{author}{Watson, D.}, \bibinfo{year}{2016}.
\newblock \bibinfo{title}{A new perspective on the ripple effect in the uk
  housing market: Comovement, cyclical subsamples and alternative indices}.
\newblock \bibinfo{journal}{Urban Studies} \bibinfo{volume}{53},
  \bibinfo{pages}{3048--3062}.
%Type = Article
\bibitem[{Curto et~al.(2015)Curto, Fregonara and Semeraro}]{curto2015listing}
\bibinfo{author}{Curto, R.}, \bibinfo{author}{Fregonara, E.},
  \bibinfo{author}{Semeraro, P.}, \bibinfo{year}{2015}.
\newblock \bibinfo{title}{Listing behaviour in the italian real estate market}.
\newblock \bibinfo{journal}{International Journal of Housing Markets and
  Analysis} .
%Type = Article
\bibitem[{Datta et~al.(2016)Datta, Banerjee, Finley and
  Gelfand}]{datta2016hierarchical}
\bibinfo{author}{Datta, A.}, \bibinfo{author}{Banerjee, S.},
  \bibinfo{author}{Finley, A.O.}, \bibinfo{author}{Gelfand, A.E.},
  \bibinfo{year}{2016}.
\newblock \bibinfo{title}{Hierarchical nearest-neighbor gaussian process models
  for large geostatistical datasets}.
\newblock \bibinfo{journal}{Journal of the American Statistical Association}
  \bibinfo{volume}{111}, \bibinfo{pages}{800--812}.
%Type = Article
\bibitem[{Deb and Tsay(2019)}]{deb2019spatio}
\bibinfo{author}{Deb, S.}, \bibinfo{author}{Tsay, R.S.}, \bibinfo{year}{2019}.
\newblock \bibinfo{title}{Spatio-temporal models with space-time interaction
  and their applications to air pollution data}.
\newblock \bibinfo{journal}{Statistica Sinica} \bibinfo{volume}{29},
  \bibinfo{pages}{1181--1207}.
%Type = Article
\bibitem[{Dubin and Sung(1990)}]{dubin1990specification}
\bibinfo{author}{Dubin, R.A.}, \bibinfo{author}{Sung, C.H.},
  \bibinfo{year}{1990}.
\newblock \bibinfo{title}{Specification of hedonic regressions: non-nested
  tests on measures of neighborhood quality}.
\newblock \bibinfo{journal}{Journal of Urban Economics} \bibinfo{volume}{27},
  \bibinfo{pages}{97--110}.
%Type = Inproceedings
\bibitem[{Feng and Jones(2016)}]{feng2016postcode}
\bibinfo{author}{Feng, Y.}, \bibinfo{author}{Jones, K.}, \bibinfo{year}{2016}.
\newblock \bibinfo{title}{Postcode or census geography? an examination of
  neighbourhood classification for house price predictions}, in:
  \bibinfo{booktitle}{The 22nd Annual Pacific Rim Real Estate Society
  Conference}, pp. \bibinfo{pages}{1--12}.
%Type = Article
\bibitem[{Finley et~al.(2019)Finley, Datta, Cook, Morton, Andersen and
  Banerjee}]{finley2019efficient}
\bibinfo{author}{Finley, A.O.}, \bibinfo{author}{Datta, A.},
  \bibinfo{author}{Cook, B.D.}, \bibinfo{author}{Morton, D.C.},
  \bibinfo{author}{Andersen, H.E.}, \bibinfo{author}{Banerjee, S.},
  \bibinfo{year}{2019}.
\newblock \bibinfo{title}{Efficient algorithms for bayesian nearest neighbor
  gaussian processes}.
\newblock \bibinfo{journal}{Journal of Computational and Graphical Statistics}
  \bibinfo{volume}{28}, \bibinfo{pages}{401--414}.
%Type = Book
\bibitem[{Fotheringham et~al.(2003)Fotheringham, Brunsdon and
  Charlton}]{fotheringham2003geographically}
\bibinfo{author}{Fotheringham, A.S.}, \bibinfo{author}{Brunsdon, C.},
  \bibinfo{author}{Charlton, M.}, \bibinfo{year}{2003}.
\newblock \bibinfo{title}{Geographically weighted regression: the analysis of
  spatially varying relationships}.
\newblock \bibinfo{publisher}{John Wiley \& Sons}.
%Type = Misc
\bibitem[{{Free map tools}(2023)}]{freemap}
\bibinfo{author}{{Free map tools}}, \bibinfo{year}{2023}.
\newblock \bibinfo{title}{{UK Postcodes with Latitude and Longitude}}.
\newblock \URLprefix
  \url{https://www.freemaptools.com/download-uk-postcode-lat-lng.htm}.
%Type = Article
\bibitem[{Fuerst et~al.(2015)Fuerst, McAllister, Nanda and
  Wyatt}]{fuerst2015does}
\bibinfo{author}{Fuerst, F.}, \bibinfo{author}{McAllister, P.},
  \bibinfo{author}{Nanda, A.}, \bibinfo{author}{Wyatt, P.},
  \bibinfo{year}{2015}.
\newblock \bibinfo{title}{Does energy efficiency matter to home-buyers? an
  investigation of epc ratings and transaction prices in england}.
\newblock \bibinfo{journal}{Energy Economics} \bibinfo{volume}{48},
  \bibinfo{pages}{145--156}.
%Type = Article
\bibitem[{Furrer et~al.(2006)Furrer, Genton and Nychka}]{furrer2006covariance}
\bibinfo{author}{Furrer, R.}, \bibinfo{author}{Genton, M.G.},
  \bibinfo{author}{Nychka, D.}, \bibinfo{year}{2006}.
\newblock \bibinfo{title}{Covariance tapering for interpolation of large
  spatial datasets}.
\newblock \bibinfo{journal}{Journal of Computational and Graphical Statistics}
  \bibinfo{volume}{15}, \bibinfo{pages}{502--523}.
%Type = Article
\bibitem[{Gelfand et~al.(2004)Gelfand, Ecker, Knight and
  Sirmans}]{gelfand2004dynamics}
\bibinfo{author}{Gelfand, A.E.}, \bibinfo{author}{Ecker, M.D.},
  \bibinfo{author}{Knight, J.R.}, \bibinfo{author}{Sirmans, C.},
  \bibinfo{year}{2004}.
\newblock \bibinfo{title}{The dynamics of location in home price}.
\newblock \bibinfo{journal}{The journal of real estate finance and economics}
  \bibinfo{volume}{29}, \bibinfo{pages}{149--166}.
%Type = Article
\bibitem[{Gerassimenko et~al.(2023)Gerassimenko, Defau and
  De~Moor}]{gerassimenko2023impact}
\bibinfo{author}{Gerassimenko, A.}, \bibinfo{author}{Defau, L.},
  \bibinfo{author}{De~Moor, L.}, \bibinfo{year}{2023}.
\newblock \bibinfo{title}{The impact of energy certificates on sales and rental
  prices: a comparative analysis}.
\newblock \bibinfo{journal}{International Journal of Housing Markets and
  Analysis} .
%Type = Techreport
\bibitem[{Geweke et~al.(1991)}]{geweke1991evaluating}
\bibinfo{author}{Geweke, J.F.}, et~al., \bibinfo{year}{1991}.
\newblock \bibinfo{title}{Evaluating the accuracy of sampling-based approaches
  to the calculation of posterior moments}.
\newblock \bibinfo{type}{Technical Report}. Federal Reserve Bank of
  Minneapolis.
%Type = Article
\bibitem[{Guhaniyogi and Banerjee(2018)}]{guhaniyogi2018meta}
\bibinfo{author}{Guhaniyogi, R.}, \bibinfo{author}{Banerjee, S.},
  \bibinfo{year}{2018}.
\newblock \bibinfo{title}{Meta-kriging: Scalable bayesian modeling and
  inference for massive spatial datasets}.
\newblock \bibinfo{journal}{Technometrics} \bibinfo{volume}{60},
  \bibinfo{pages}{430--444}.
%Type = Article
\bibitem[{Guhaniyogi et~al.(2022)Guhaniyogi, Li, Savitsky and
  Srivastava}]{guhaniyogi2022distributed}
\bibinfo{author}{Guhaniyogi, R.}, \bibinfo{author}{Li, C.},
  \bibinfo{author}{Savitsky, T.}, \bibinfo{author}{Srivastava, S.},
  \bibinfo{year}{2022}.
\newblock \bibinfo{title}{Distributed bayesian inference in massive spatial
  data}.
\newblock \bibinfo{journal}{Statist. Sci} .
%Type = Article
\bibitem[{Hijmans(2021)}]{hijmans2021introduction}
\bibinfo{author}{Hijmans, R.J.}, \bibinfo{year}{2021}.
\newblock \bibinfo{title}{Introduction to the ``geosphere” package (version
  1.5-14)} .
%Type = Article
\bibitem[{Holly et~al.(2010)Holly, Pesaran and Yamagata}]{holly2010spatio}
\bibinfo{author}{Holly, S.}, \bibinfo{author}{Pesaran, M.H.},
  \bibinfo{author}{Yamagata, T.}, \bibinfo{year}{2010}.
\newblock \bibinfo{title}{{A spatio-temporal model of house prices in the
  USA}}.
\newblock \bibinfo{journal}{Journal of Econometrics} \bibinfo{volume}{158},
  \bibinfo{pages}{160--173}.
%Type = Article
\bibitem[{Hyndman and Khandakar(2008)}]{hyndman2008automatic}
\bibinfo{author}{Hyndman, R.J.}, \bibinfo{author}{Khandakar, Y.},
  \bibinfo{year}{2008}.
\newblock \bibinfo{title}{Automatic time series forecasting: the forecast
  package for r}.
\newblock \bibinfo{journal}{Journal of statistical software}
  \bibinfo{volume}{27}, \bibinfo{pages}{1--22}.
%Type = Article
\bibitem[{Li et~al.(2017)Li, Srivastava and Dunson}]{li2017simple}
\bibinfo{author}{Li, C.}, \bibinfo{author}{Srivastava, S.},
  \bibinfo{author}{Dunson, D.B.}, \bibinfo{year}{2017}.
\newblock \bibinfo{title}{Simple, scalable and accurate posterior interval
  estimation}.
\newblock \bibinfo{journal}{Biometrika} \bibinfo{volume}{104},
  \bibinfo{pages}{665--680}.
%Type = Article
\bibitem[{Liu(2013)}]{liu2013spatial}
\bibinfo{author}{Liu, X.}, \bibinfo{year}{2013}.
\newblock \bibinfo{title}{Spatial and temporal dependence in house price
  prediction}.
\newblock \bibinfo{journal}{The Journal of Real Estate Finance and Economics}
  \bibinfo{volume}{47}, \bibinfo{pages}{341--369}.
%Type = Inproceedings
\bibitem[{Ma and Li(2017)}]{ma2017impacts}
\bibinfo{author}{Ma, H.}, \bibinfo{author}{Li, J.}, \bibinfo{year}{2017}.
\newblock \bibinfo{title}{The impacts of supply and demand analysis on the
  price of the real estate market}, in: \bibinfo{booktitle}{7th International
  Conference on Education, Management, Information and Mechanical Engineering
  (EMIM 2017)}, \bibinfo{organization}{Atlantis Press}. pp.
  \bibinfo{pages}{1881--1885}.
%Type = Article
\bibitem[{Mete and Yomralioglu(2022)}]{mete2022hybrid}
\bibinfo{author}{Mete, M.O.}, \bibinfo{author}{Yomralioglu, T.},
  \bibinfo{year}{2022}.
\newblock \bibinfo{title}{A hybrid approach for mass valuation of residential
  properties through geographic information systems and machine learning
  integration}.
\newblock \bibinfo{journal}{Geographical Analysis} .
%Type = Inproceedings
\bibitem[{Minsker et~al.(2014)Minsker, Srivastava, Lin and
  Dunson}]{minsker2014scalable}
\bibinfo{author}{Minsker, S.}, \bibinfo{author}{Srivastava, S.},
  \bibinfo{author}{Lin, L.}, \bibinfo{author}{Dunson, D.},
  \bibinfo{year}{2014}.
\newblock \bibinfo{title}{Scalable and robust bayesian inference via the median
  posterior}, in: \bibinfo{booktitle}{International conference on machine
  learning}, \bibinfo{organization}{PMLR}. pp. \bibinfo{pages}{1656--1664}.
%Type = Article
\bibitem[{Moran(1950)}]{moran1950notes}
\bibinfo{author}{Moran, P.A.}, \bibinfo{year}{1950}.
\newblock \bibinfo{title}{Notes on continuous stochastic phenomena}.
\newblock \bibinfo{journal}{Biometrika} \bibinfo{volume}{37},
  \bibinfo{pages}{17--23}.
%Type = Article
\bibitem[{Neal(2003)}]{neal2003slice}
\bibinfo{author}{Neal, R.M.}, \bibinfo{year}{2003}.
\newblock \bibinfo{title}{Slice sampling}.
\newblock \bibinfo{journal}{The annals of statistics} \bibinfo{volume}{31},
  \bibinfo{pages}{705--767}.
%Type = Article
\bibitem[{Nemeth and Fearnhead(2021)}]{nemeth2021stochastic}
\bibinfo{author}{Nemeth, C.}, \bibinfo{author}{Fearnhead, P.},
  \bibinfo{year}{2021}.
\newblock \bibinfo{title}{Stochastic gradient markov chain monte carlo}.
\newblock \bibinfo{journal}{Journal of the American Statistical Association}
  \bibinfo{volume}{116}, \bibinfo{pages}{433--450}.
%Type = Misc
\bibitem[{{Office for national statistics}(2018)}]{ons}
\bibinfo{author}{{Office for national statistics}}, \bibinfo{year}{2018}.
\newblock \bibinfo{title}{{Exploring recent trends in the London housing
  market}}.
\newblock \URLprefix
  \url{https://www.ons.gov.uk/economy/inflationandpriceindices/articles/exploringrecenttrendsinthelondonhousingmarket/2018-09-19}.
%Type = Article
\bibitem[{Ou et~al.(2021)Ou, Sen and Dunson}]{ou2021scalable}
\bibinfo{author}{Ou, R.}, \bibinfo{author}{Sen, D.}, \bibinfo{author}{Dunson,
  D.}, \bibinfo{year}{2021}.
\newblock \bibinfo{title}{Scalable bayesian inference for time series via
  divide-and-conquer}.
\newblock \bibinfo{journal}{arXiv preprint arXiv:2106.11043} .
%Type = Article
\bibitem[{Pace et~al.(1998)Pace, Barry, Clapp and
  Rodriquez}]{pace1998spatiotemporal}
\bibinfo{author}{Pace, R.K.}, \bibinfo{author}{Barry, R.},
  \bibinfo{author}{Clapp, J.M.}, \bibinfo{author}{Rodriquez, M.},
  \bibinfo{year}{1998}.
\newblock \bibinfo{title}{Spatiotemporal autoregressive models of neighborhood
  effects}.
\newblock \bibinfo{journal}{The Journal of Real Estate Finance and Economics}
  \bibinfo{volume}{17}, \bibinfo{pages}{15--33}.
%Type = Article
\bibitem[{Pace and Gilley(1998)}]{pace1998generalizing}
\bibinfo{author}{Pace, R.K.}, \bibinfo{author}{Gilley, O.W.},
  \bibinfo{year}{1998}.
\newblock \bibinfo{title}{Generalizing the ols and grid estimators}.
\newblock \bibinfo{journal}{Real Estate Economics} \bibinfo{volume}{26},
  \bibinfo{pages}{331--347}.
%Type = Article
\bibitem[{Paradis and Schliep(2019)}]{Schliep2019}
\bibinfo{author}{Paradis, E.}, \bibinfo{author}{Schliep, K.},
  \bibinfo{year}{2019}.
\newblock \bibinfo{title}{ape 5.0: an environment for modern phylogenetics and
  evolutionary analyses in {R}}.
\newblock \bibinfo{journal}{Bioinformatics} \bibinfo{volume}{35},
  \bibinfo{pages}{526--528}.
\newblock \DOIprefix\doi{10.1093/bioinformatics/bty633}.
%Type = Article
\bibitem[{Quiroz et~al.(2019)Quiroz, Kohn, Villani and
  Tran}]{quiroz2019speeding}
\bibinfo{author}{Quiroz, M.}, \bibinfo{author}{Kohn, R.},
  \bibinfo{author}{Villani, M.}, \bibinfo{author}{Tran, M.N.},
  \bibinfo{year}{2019}.
\newblock \bibinfo{title}{Speeding up mcmc by efficient data subsampling}.
\newblock \bibinfo{journal}{Journal of the American Statistical Association}
  \bibinfo{volume}{114}, \bibinfo{pages}{831--843}.
%Type = Article
\bibitem[{Rosen(1974)}]{rosen1974hedonic}
\bibinfo{author}{Rosen, S.}, \bibinfo{year}{1974}.
\newblock \bibinfo{title}{Hedonic prices and implicit markets: product
  differentiation in pure competition}.
\newblock \bibinfo{journal}{Journal of political economy} \bibinfo{volume}{82},
  \bibinfo{pages}{34--55}.
%Type = Book
\bibitem[{Sahu(2022)}]{sahu2022bayesian}
\bibinfo{author}{Sahu, S.}, \bibinfo{year}{2022}.
\newblock \bibinfo{title}{Bayesian modeling of spatio-temporal data with R}.
\newblock \bibinfo{publisher}{CRC Press}.
%Type = Article
\bibitem[{Sahu et~al.(2006)Sahu, Gelfand and Holland}]{sahu2006spatio}
\bibinfo{author}{Sahu, S.K.}, \bibinfo{author}{Gelfand, A.E.},
  \bibinfo{author}{Holland, D.M.}, \bibinfo{year}{2006}.
\newblock \bibinfo{title}{Spatio-temporal modeling of fine particulate matter}.
\newblock \bibinfo{journal}{Journal of Agricultural, Biological, and
  Environmental Statistics} \bibinfo{volume}{11}, \bibinfo{pages}{61--86}.
%Type = Article
\bibitem[{Sahu et~al.(2010)Sahu, Gelfand and Holland}]{sahu2010fusing}
\bibinfo{author}{Sahu, S.K.}, \bibinfo{author}{Gelfand, A.E.},
  \bibinfo{author}{Holland, D.M.}, \bibinfo{year}{2010}.
\newblock \bibinfo{title}{Fusing point and areal level space--time data with
  application to wet deposition}.
\newblock \bibinfo{journal}{Journal of the Royal Statistical Society: Series C
  (Applied Statistics)} \bibinfo{volume}{59}, \bibinfo{pages}{77--103}.
%Type = Article
\bibitem[{Shyamalkumar and Srivastava(2022)}]{shyamalkumar2022algorithm}
\bibinfo{author}{Shyamalkumar, N.D.}, \bibinfo{author}{Srivastava, S.},
  \bibinfo{year}{2022}.
\newblock \bibinfo{title}{An algorithm for distributed bayesian inference}.
\newblock \bibinfo{journal}{Stat} \bibinfo{volume}{11}, \bibinfo{pages}{e432}.
%Type = Article
\bibitem[{Soltani et~al.(2021)Soltani, Pettit, Heydari and
  Aghaei}]{soltani2021housing}
\bibinfo{author}{Soltani, A.}, \bibinfo{author}{Pettit, C.J.},
  \bibinfo{author}{Heydari, M.}, \bibinfo{author}{Aghaei, F.},
  \bibinfo{year}{2021}.
\newblock \bibinfo{title}{Housing price variations using spatio-temporal data
  mining techniques}.
\newblock \bibinfo{journal}{Journal of Housing and the Built Environment} ,
  \bibinfo{pages}{1--29}.
%Type = Article
\bibitem[{Srivastava et~al.(2018)Srivastava, Li and
  Dunson}]{srivastava2018scalable}
\bibinfo{author}{Srivastava, S.}, \bibinfo{author}{Li, C.},
  \bibinfo{author}{Dunson, D.B.}, \bibinfo{year}{2018}.
\newblock \bibinfo{title}{Scalable bayes via barycenter in wasserstein space}.
\newblock \bibinfo{journal}{The Journal of Machine Learning Research}
  \bibinfo{volume}{19}, \bibinfo{pages}{312--346}.
%Type = Article
\bibitem[{Szab{\'o} et~al.(2019)Szab{\'o}, Van~Zanten
  et~al.}]{szabo2019asymptotic}
\bibinfo{author}{Szab{\'o}, B.}, \bibinfo{author}{Van~Zanten, H.}, et~al.,
  \bibinfo{year}{2019}.
\newblock \bibinfo{title}{An asymptotic analysis of distributed nonparametric
  methods}.
\newblock \bibinfo{journal}{Journal of Machine Learning Research}
  \bibinfo{volume}{20}, \bibinfo{pages}{1--30}.
%Type = Article
\bibitem[{Teye and Ahelegbey(2017)}]{teye2017detecting}
\bibinfo{author}{Teye, A.L.}, \bibinfo{author}{Ahelegbey, D.F.},
  \bibinfo{year}{2017}.
\newblock \bibinfo{title}{{Detecting spatial and temporal house price diffusion
  in the Netherlands: A Bayesian network approach}}.
\newblock \bibinfo{journal}{Regional Science and Urban Economics}
  \bibinfo{volume}{65}, \bibinfo{pages}{56--64}.
%Type = Article
\bibitem[{Wikle(2010)}]{wikle2010low}
\bibinfo{author}{Wikle, C.K.}, \bibinfo{year}{2010}.
\newblock \bibinfo{title}{Low-rank representations for spatial processes}.
\newblock \bibinfo{journal}{Handbook of spatial statistics}
  \bibinfo{volume}{107}, \bibinfo{pages}{118}.
%Type = Book
\bibitem[{Zhang(2006)}]{zhang2006schur}
\bibinfo{author}{Zhang, F.}, \bibinfo{year}{2006}.
\newblock \bibinfo{title}{The Schur complement and its applications}.
  volume~\bibinfo{volume}{4}.
\newblock \bibinfo{publisher}{Springer Science \& Business Media}.
%Type = Article
\bibitem[{Zhang(2016)}]{zhang2016flood}
\bibinfo{author}{Zhang, L.}, \bibinfo{year}{2016}.
\newblock \bibinfo{title}{Flood hazards impact on neighborhood house prices: A
  spatial quantile regression analysis}.
\newblock \bibinfo{journal}{Regional Science and Urban Economics}
  \bibinfo{volume}{60}, \bibinfo{pages}{12--19}.

\end{thebibliography}

\end{document}